\documentclass[twocolumn]{aastex631}

\usepackage[T1]{fontenc}
\usepackage{xspace}
\usepackage{multirow}
\usepackage{amsmath}
\usepackage{lipsum}
\usepackage{soul}
\usepackage[normalem]{ulem}
\usepackage{float}
\usepackage{comment}

\newcommand{\E}[1]{\ensuremath{\times 10^{#1}} }

\newcommand{\rxte}{\textit{RXTE}\xspace}

\newcommand{\nicer}{\textit{NICER}\xspace}

\newcommand{\source}{4U~1820--30\xspace}

\newcommand{\fa}{$f_a$~}

\shorttitle{X-ray bursts from 4U 1820--30 with \nicer}
\shortauthors{Jaisawal et al.}
\graphicspath{{./}{figures/}}
\begin{document}

\title{A Comprehensive Study of Thermonuclear X-ray Bursts from 4U 1820--30 with \nicer: Accretion Disk Interactions and a Candidate Burst Oscillation}
\author[0000-0002-6789-2723]{Gaurava K. Jaisawal}\email{gaurava@space.dtu.dk}
\affiliation{DTU Space, Technical University of Denmark, 
  Elektrovej 327-328, DK-2800 Lyngby, Denmark}

\author[0000-0002-5665-3452]{Z. Funda Bostanc\i~}
\affiliation{Istanbul University, Science Faculty, Department of Astronomy and Space Sciences, Beyaz\i t, 34119, \.Istanbul, T\"urkiye}
\affiliation{Istanbul University Observatory Research and Application Center, Istanbul University 34119, \.Istanbul T\"urkiye}

\author[0000-0002-4729-1592]{Tu\u{g}ba Boztepe}
\affiliation{Istanbul University, Graduate School of Sciences, Department of Astronomy and Space Sciences, Beyaz\i t, 34119, \.Istanbul, T\"urkiye}

\author[0000-0002-3531-9842]{Tolga G\"uver}
\affiliation{Istanbul University, Science Faculty, Department of Astronomy and Space Sciences, Beyaz\i t, 34119, \.Istanbul, T\"urkiye}
\affiliation{Istanbul University Observatory Research and Application Center, Istanbul University 34119, \.Istanbul T\"urkiye}

 \author[0000-0001-7681-5845]{Tod E. Strohmayer} 
 \affil{Astrophysics Science Division and Joint Space-Science Institute,
  NASA's Goddard Space Flight Center, Greenbelt, MD 20771, USA}

\author[0000-0001-8128-6976]{David R. Ballantyne}
\affiliation{Center for Relativistic Astrophysics, School of Physics, Georgia Institute of Technology, 837 State Street, Atlanta, GA 30332-0430, USA}

\author{Jens H. Beck}
\affiliation{DTU Space, Technical University of Denmark, 
  Elektrovej 327-328, DK-2800 Lyngby, Denmark}

\author[0000-0002-5274-6790]{Ersin G\"o\u{g}\"u\c{s}}
\affiliation{Faculty of Engineering and Natural Sciences, Sabanc\i~University, Orhanl\i-Tuzla 34956, \.Istanbul, T\"urkiye }

\author[0000-0002-3422-0074]{Diego Altamirano}
\affiliation{School of Physics and Astronomy, University of Southampton, Southampton SO17 1BJ, UK}

 \author{Zaven Arzoumanian} 
 \affiliation{Astrophysics Science Division, 
   NASA's Goddard Space Flight Center, Greenbelt, MD 20771, USA}

\author[0000-0001-8804-8946]{Deepto Chakrabarty}
 \affil{MIT Kavli Institute for Astrophysics and Space Research, 
   Massachusetts Institute of Technology, Cambridge, MA 02139, USA}
   
 \author[0000-0001-7115-2819]{Keith C. Gendreau} 
 \affiliation{Astrophysics Science Division, 
   NASA's Goddard Space Flight Center, Greenbelt, MD 20771, USA}

\author[0000-0002-6449-106X]{Sebastien~Guillot}
\affil{IRAP, CNRS, 9 avenue du Colonel Roche, BP 44346, F-31028 Toulouse Cedex 4, France}
\affil{Universit\'{e} de Toulouse, CNES, UPS-OMP, F-31028 Toulouse, France.}

\author[0000-0002-8961-939X]{Renee M. Ludlam}
\affiliation{Department of Physics \& Astronomy, Wayne State University, 666 West Hancock Street, Detroit, MI 48201, USA}

\author[0000-0002-0940-6563]{Mason Ng}
\affiliation{MIT Kavli Institute for Astrophysics and Space Research, Massachusetts Institute of Technology, Cambridge, MA 02139, USA}

\author[0000-0002-0118-2649]{Andrea Sanna}
\affiliation{Dipartimento di Fisica, Universit\`a degli Studi di Cagliari, SP Monserrato-Sestu km 0.7, I-09042 Monserrato, Italy}

\author[0000-0002-4397-8370]{J{\'e}r{\^o}me Chenevez}
\affiliation{DTU Space, Technical University of Denmark, 
  Elektrovej 327-328, DK-2800 Lyngby, Denmark}

\begin{abstract}

We present the results obtained from timing and spectral studies of 15 thermonuclear X-ray bursts from \source observed with the Neutron Star Interior Composition Explorer (\nicer) during its five years of observations between 2017-2022. All bursts showed clear signs of photospheric radius expansion, where the neutron star (NS) photosphere expanded more than 50~km above the surface. One of the bursts produced a super-expansion with a blackbody emission radius of 902~km for the first time with \nicer. We searched for burst oscillations in all 15 bursts and found evidence of a coherent oscillation at 716~Hz in a burst, with a 2.9$\sigma$ detection level based on Monte Carlo simulations. If confirmed with future observations, \source would become the fastest-spinning NS known in X-ray binary systems.  The fractional rms amplitude of the candidate burst oscillation was found to be 5.8\% in the energy range of 3--10~keV. Following the variable persistent model from burst time-resolved spectroscopy, an anti-correlation is seen between the maximum scaling factor value and the (pre-burst) persistent flux. We detected a low value of ionization at the peak of each burst based on reflection modeling of burst spectra. A partially interacting inner accretion disk or a weakly ionized outer disk may cause the observed ionization dip during the photospheric radius expansion phase. 

\end{abstract}

 
\section{Introduction} \label{sec:intro}

Thermonuclear X-ray bursts are observed from weakly magnetized neutron stars (NSs) in low-mass X-ray binaries (LMXBs). In these systems, accreted material such as hydrogen and/ or helium from a donor star undergoes unstable thermonuclear burning on the neutron star's surface.  These bursts typically show a fast rise for a few seconds, followed by an exponential decay for tens of seconds. The decay profile represents the cooling of the burnt fuel layer on the surface. The burst emission can usually be characterized by a simple black-body spectrum with a temperature varying between 1--3 keV. Depending on the accretion rate and spectral state of the neutron star, X-ray bursts can occur quasi-periodically from hours to days timescales (refer to \citealt{Lewin1993SSRv...62..223L, Strohmayer2006csxs.book..113S, 2021ASSL..461..209G} for X-ray bursts review).

\source is an ultra-compact X-ray binary (UCXB) located in a metal-rich globular cluster NGC~6624 at a distance of 8.4 kpc  \citep{Grindlay1976ApJ...205L.127G, Stella1987ApJ...312L..17S, Valenti2004MNRAS.351.1204V}. It has the shortest known orbital period of 11.4 minutes among neutron star LMXBs \citep{Stella1987ApJ...312L..17S}. UCXBs are a rare subset of LMXBs that harbor a neutron star or black hole as a compact object with an orbital period of less than an hour \citep{Nelson1986ApJ...304..231N}. Such a short orbital period indicates a relatively small size of the binary system, which cannot fit a main-sequence companion star in its Roche lobe. The companion (donor) star is thus believed to be a degenerate white dwarf or a helium star that has been stripped of its outer layers \citep{Rappaport1987ApJ...322..842R}. As a result, the thermonuclear bursts from \source are considered to originate from helium-rich, but hydrogen-deficient fuel  \citep{Cumming2003ApJ...595.1077C}.

The persistent emission from LMXBs is dominated by a softer or harder spectral component depending on the source accretion state (see, e.g., \citealt{Done2007A&ARv..15....1D}). In a high-soft state, soft X-ray photons from the accretion disk contribute majorly to the accretion emission at higher luminosities. Whereas, a non-thermal power-law component dominates the emission at lower luminosities, corresponding to a low-hard state of a source.
The persistent X-ray emission from \source is usually observed in a high-soft state with a luminosity reaching up to 50\% of the Eddington luminosity \citep{Titarchuk2013ApJ...767..160T}.  
But it temporarily transitions to a low accretion state for a few days with a drop of two to three-fold in X-ray flux  \citep{Priedhorsky1984ApJ...284L..17P, Farrell2009MNRAS.393..139F}. This high-to-low state transition occurs quasi-periodically at about 176 days \citep{Strohmayer2002ApJ...566.1045S}. Only in this low-hard state, strong thermonuclear X-ray bursts are observed from \source, at a recurrence time of 2--4 hours  \citep{Grindlay1976ApJ...205L.127G, Chou2001ApJ...563..934C, Galloway2008ApJS..179..360G, Zand2012A&A...547A..47I}. The bursts are not detected during the high-soft state, possibly due to stable thermonuclear burning of the accreted material, which cannot be easily distinguished from intense accretion emission (see, e.g., \citealt{Bildsten1995ApJ...438..852B, Zand2012A&A...547A..47I}).

\source typically displays short X-ray bursts, only lasting for 10--15 seconds. This is likely because the ignited helium-rich fuel burns out rapidly on the surface via the triple-$\alpha$ process, resulting in short but powerful photospheric radius expansion (PRE) bursts \citep{Cumming2003ApJ...595.1077C}.
The PRE phenomenon occurs at the Eddington luminosity, where the neutron star photosphere expands suddenly due to immense radiation pressure. The expansion is accompanied by a drop in the photospheric temperature near the burst peak at a constant luminosity level \citep{Kuulkers2003}. Sometimes, this effect is visible in the hard X-ray light curve above 3~keV in the form of a double-peaked profile. The observed behavior appears since the photospheric temperature decreases significantly at the burst peak and lies beyond the detection band-pass of hard X-ray instruments. Thermonuclear bursts have been used to measure the mass and the radius of the neutron star in \source \citep{2010ApJ...719.1807G,2016ApJ...820...28O, Suleimanov2017MNRAS.472.3905S}, but see also \citet{2013MNRAS.429.3266G} and \citet{Kajava2014MNRAS.445.4218K}.
In addition to short bursts, \source is known to produce more energetic superbursts that last several hours. The superbursts are powered by the burning of carbon layer beneath the surface  \citep{Strohmayer2002ApJ...566.1045S, Ballantyne2004ApJ...602L.105B}.

\begin{figure*}[ht!]
\centering
\includegraphics[height=5.3in, width=7in, angle=0]{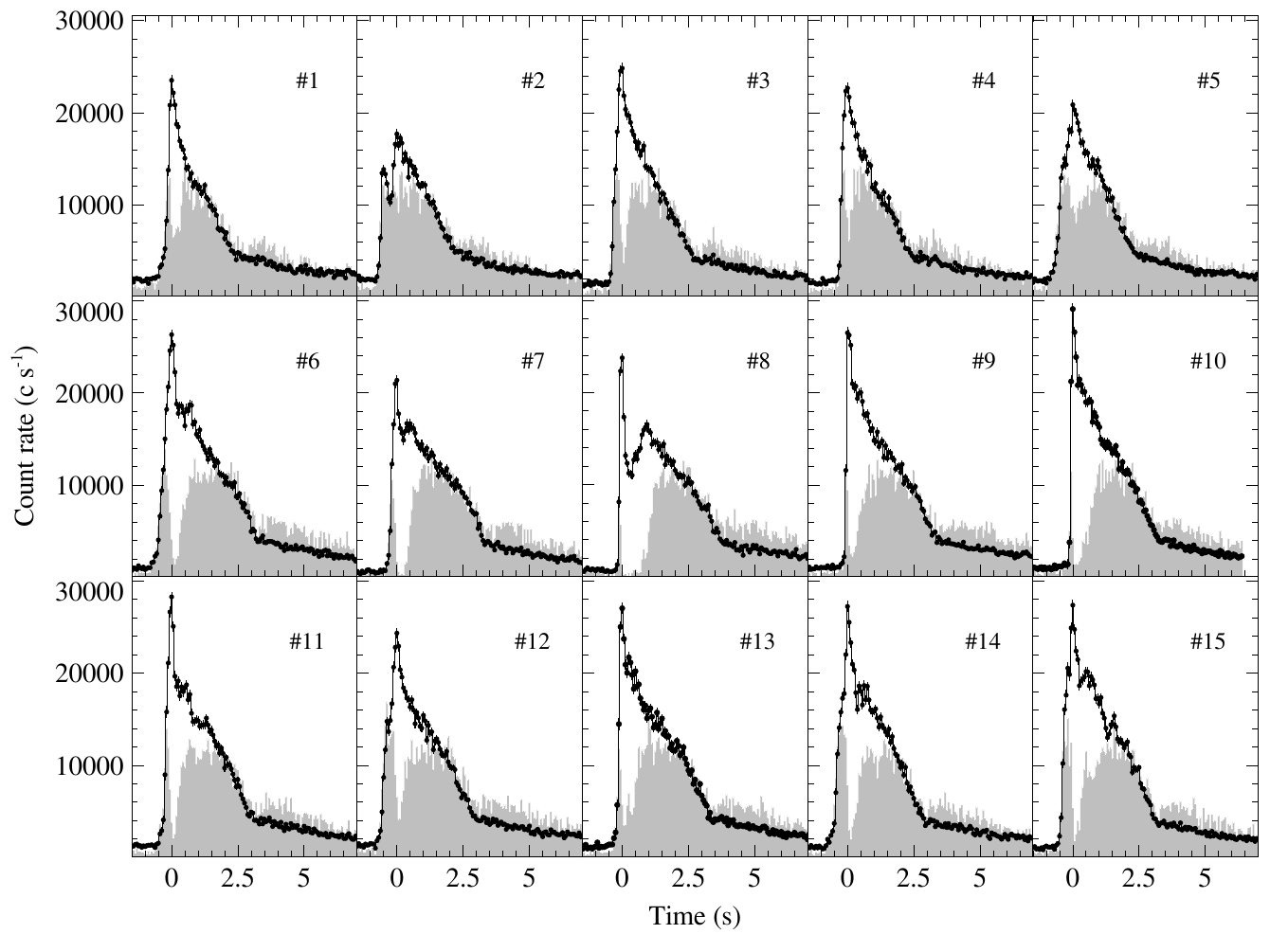}
\caption{Thermonuclear X-ray bursts from \source at 1/16~s time resolution in the 0.4--12 keV range observed with \nicer. The grey shaded area corresponds to the burst profile in the 3--12 keV range that is scaled up by a factor of five for plotting purposes. All the bursts are aligned on a relative time scale to keep peak time at zero.} 
\label{burst_lc}
\end{figure*}

X-ray photons from powerful thermonuclear bursts can affect the accretion disk environment, particularly around the burst peak \citep{Degenaar2018}. This has been observed as deviations from pure black-body emission of X-ray bursts from several LMXBs such as Aql X-1 \citep{Keek2018ApJ...855L...4K, Guver2022MNRAS.510.1577G}, \source \citep{Keek2018ApJ...856L..37K}, 4U 1608-52 \citep{Jaisawal2019ApJ...883...61J, 2021ApJ...910...37G}, and 4U~1636$-$536 \citep{Chen2018ApJ...864L..30C, Guver2022ApJ...935..154G}. This phenomenon can be understood in terms of reprocessing/reflection from the accretion disk \citep{Ballantyne2004MNRAS.351...57B, Speicher2022MNRAS.509.1736S}, an evolving accretion flow rate due to Poynting-Robertson drag \citep{Walker1992, Speicher2023MNRAS.526.1388S}, or through cooling of the Compton corona \citep{Ji2014a, Speicher2020MNRAS.499.4479S}. 
Additionally, as an indication of burst-disk interaction and burst-driven winds, a \nicer study of \source found evidence of spectral features at 1.0 (emission), 1.7, and 3.0 keV (both in absorption) from two pairs of PRE bursts observed in August 2017 \citep{Strohmayer2019ApJ...878L..27S}.  A systematic shift between the lines was detected by comparing co-added burst spectra. The observed shifts were explained by considering a possible combination of gravitational and Doppler effects on the spectral features originating from burst-driven winds of the PRE bursts. 

Burst oscillations observed during some X-ray bursts are intriguing phenomena that provide valuable insights into the neutron star interior and surface physics \citep{Strohmayer2006csxs.book..113S, A.Watts2012, Bhattacharyya2022ASSL..465..125B}. These nuclear-powered oscillations coincide with the spin frequency of the neutron star \citep{Chakrabarty2003Natur.424...42C}, and are firmly detected in a range of 11 to 620~Hz from multiple sources such as weakly magnetized accretion-powered X-ray pulsars and X-ray bursters in LMXBs \citep{A.Watts2012, Bhattacharyya2022ASSL..465..125B}. 

No definite burst oscillation has been reported from \source, although an oscillation candidate only using a 1~s time window was found around 280 Hz in \rxte data \citep{Bilous2019}. On the other hand, the source is known to show twin kHz quasi-periodic oscillations (QPOs) from the persistent emission, in lower and upper banana branches as well as in the island state, with a frequency difference of about 275~Hz \citep{Smale1997ApJ...483L.119S, Zhang1998ApJ...500L.171Z, Altamirano2005ApJ...633..358A}.  Apart from high-frequency QPOs, 1~mHz QPO, along with its first harmonics, were detected in high and low luminosity states of \source \citep{Chen2023MNRAS.523.2663C}. In the absence of any substantial drift, the 1~mHz QPO is explained in terms of the free precession period of the NS in the system rather than marginally stable burning of the material \citep{Chen2023MNRAS.523.2663C}.

In this paper, we study a sample of 15 thermonuclear X-ray bursts from \source using \nicer data collected between June 2017 and May 2022. 
We performed a dedicated timing analysis of these bursts that resulted in the detection of a candidate burst oscillation at 716 Hz. We investigated the spectral evolution of the bursts and their effect on the accretion environment based on time-resolved spectroscopy using different models. \citet{Yu2023arXiv231216420Y} has also studied these bursts with the same \nicer data using various spectral models such as a variable persistent emission method and a disk reflection model {\tt relxillNS} model. Spectral evolution of burst parameters suggests that the bursts are of PRE nature \citep{Yu2023arXiv231216420Y}. 
In our study, we also perform burst spectral studies using the variable persistent flux method but additionally apply two independent methods that allow us to examine the effect of strong bursts on accretion emission, as presented in Sections~4.2 and 4.3. The results from these models are compared using the flux-temperature diagrams in Section~4.4.  
We further present additional findings by exploring a connection between pre-burst accretion and burst parameters in the discussion section, in addition to in-depth timing analysis in Section~3.3.
In general, our observation and data analysis are presented in Section~2. The results from timing analysis, including searches for burst oscillations, are presented in Section~\ref{timing-sec}. Our spectral findings, followed by a discussion and conclusions, are summarized in Sections~\ref{spec-sec}, \ref{sec:dis}, and ~\ref{sec:conclusion}, respectively.

\begin{table*}[]
\centering
\caption{Overview of X-ray bursts from \source observed with \nicer}
\begin{tabular}{cccccccccc}
\hline 
Burst   &ObsID  &Onset date   &Burst peak  &Pre-burst    &$\Delta$t$^b$ &Burst peak   &Fluence$^d$ &Characteristic$^e$ \\
Number   &         &(MJD) &rate$^a$ (c~s$^{-1}$) &rate (c~s$^{-1}$) &(s) &flux$^c$        &($\times$10$^{-7}$~erg~cm$^{-2}$) &time scale (s)\\
\hline
1   &1050300108 &57994.37034    &21712$\pm$613  &1808$\pm$4 &7.7    &8.3$\pm$2.2   &2.52$\pm$0.3   &3.1$\pm$1.0\\
2   &1050300108 &57994.46090    &15936$\pm$532  &1760$\pm$5 &7.9    &8.4$\pm$2.1   &2.59$\pm$0.31  &3.1$\pm$0.9\\
3   &1050300109 &57995.22207    &23376$\pm$630  &1456$\pm$6 &10.4   &8.2$\pm$0.5    &2.37$\pm$0.19  &2.9$\pm$0.3\\
4   &1050300109 &57995.33811    &21088$\pm$602  &1584$\pm$6 &8.1    &8.3$\pm$0.4    &2.27$\pm$0.26  &2.8$\pm$0.4\\
5   &1050300109 &57995.60251    &19184$\pm$578  &1680$\pm$8 &9      &9.0$\pm$2.4    &2.59$\pm$0.32  &2.9$\pm$0.9\\
6   &2050300104 &58641.28949    &25328$\pm$649  &992$\pm$7  &15.2   &8.4$\pm$2.0    &2.98$\pm$0.24  &3.6$\pm$1\\
7   &2050300108 &58646.83376    &20768$\pm$584  &608$\pm$5  &20.4   &7.0$\pm$1.5    &2.43$\pm$0.19  &3.5$\pm$0.9\\
8   &2050300110 &58648.32026    &23152$\pm$617  &640$\pm$6  &20.6   &9.5$\pm$2.7    &3.10$\pm$0.32  &3.3$\pm$1.1\\
9   &2050300115 &58655.08555    &25456$\pm$651  &1056$\pm$8 &13.7   &7.6$\pm$2.3    &2.93$\pm$0.24  &3.9$\pm$1.5\\
10  &2050300119 &58660.30662    &28144$\pm$682  &960$\pm$6  &6.8    &7.7$\pm$0.5    &2.01$\pm$0.18  &2.6$\pm$0.3\\
11  &2050300119 &58660.76490    &27088$\pm$672  &1184$\pm$5 &9.8    &8.0$\pm$0.5    &2.77$\pm$0.20  &3.5$\pm$0.3\\
12  &2050300120 &58661.77954    &23072$\pm$624  &1264$\pm$7 &11.6   &11.4$\pm$5.2    &4.07$\pm$0.48  &3.6$\pm$2\\
13  &2050300122 &58663.97779    &25872$\pm$657  &1168$\pm$6 &13.1   &7.5$\pm$0.5    &2.77$\pm$0.19  &3.7$\pm$0.4\\
14  &2050300124 &58665.58749    &26128$\pm$660  &1104$\pm$6 &10.8   &8.3$\pm$0.6    &2.78$\pm$0.24  &3.3$\pm$0.4\\
15  &4680010101 &59336.60807    &26400$\pm$662  &976$\pm$5  &12     &8.1$\pm$0.5    &2.54$\pm$0.21  &3.1$\pm$0.3\\
\hline
\end{tabular}
\label{tab:bursts}\\
\footnotesize{$^a$ Pre-burst count rates are subtracted from the observed burst peak rate in the 0.4-12~keV light curve. }\\
\footnotesize{$^b$ Burst duration based on a threshold level above the persistent count rate.}\\
\footnotesize{$^c$ The burst peak flux in  10$^{-8}$~erg~cm$^{-2}$~s$^{-1}$ unit from time-resolved burst spectroscopy (Section~4.2). }\\
\footnotesize{$^d$  Bolometric fluence of bursts estimated as a total sum of blackbody flux integrated over each time interval from time-resolved spectroscopy, using the variable persistent model (see Section~4.2).}\\
\footnotesize{$^e$ Burst characteristic time scale is defined as the ratio of the bolometric fluence to peak bolometric flux.}\\
\end{table*}

 \section{Observations and Data analysis} \label{sec:obs}

The \nicer X-ray Timing Instrument (XTI, \citealt{Gendreau2012}) is a non-imaging soft X-ray telescope attached to the International Space  Station. It consists of 56 co-aligned concentrator optics, each paired with a silicon-drift detector  \citep{Prigozhin2012}. The instrument records X-ray photons between 0.2--12~keV at unprecedented timing and good spectral resolutions of $\approx$100~ns and $\approx$100~eV, respectively. The peak effective area of the 52  currently active detectors is $\approx$1900~cm$^2$ at $1.5$~keV.

\nicer has monitored  \source actively since its launch in June 2017. In the first five years (June 2017- May 2022), a total exposure of 717~ks was devoted to the source across 170 observation ids (ObsIDs) under series of 0503001xx, 10503001xx, 20503001xx, 30503001xx, 40503001xx, 46630101xx, 46800101xx,  50503001xx, and 56040101xx. We processed these data with the standard {\tt nicerl2} pipeline using {\tt HEASoft}~v6.30. The following filtering constraints along with {\tt nimaketime}\footnote{\url{https://heasarc.gsfc.nasa.gov/lheasoft/ftools/headas/nimaketime.html}} criteria were considered:  elevation angle from the Earth limb $\texttt{ELV}>15^\circ$; bright Earth limb angle $\texttt{BR\_EARTH}>30^\circ$, source angular offset of $\texttt{ANG\_DIST}<0\fdg015$; undershoot rate range of $\texttt{underonly\_range}=$~0--400, overshoot rate range of $\texttt{overonly\_range}=$~0--2. 

We studied uncleaned ({\tt ufa}) and clean ({\tt cl}) events data files of \source with \nicer to search for the bursts in these data sets. A total of 15 thermonuclear X-ray bursts were found from 11 ObsIDs, mainly in August 2017 (1050300108 \& 1050300109), June 2019 (2050300104, 2050300108, 2050300110, 2050300115, 2050300119, 2050300120, \& 2050300122), July 2019 (2050300124), and May 2021 (4680010101). Among these, the above filtering criteria allowed the detection of 14 thermonuclear bursts in cleaned event data that contains cleaned and calibrated science events. One burst under ObsID 2050300120 at the onset time of MJD 58661.77953684, i.e., Burst~\#12 was filtered out from clean event data. We retained this burst manually in our analysis.  We note that \citet{Yu2023arXiv231216420Y} found only 12 bursts in clean event files and the remaining 3 bursts (Burst~\#9, \#12, and \#13) in ufa files due to their filtering criteria. 

A detailed log on the observed bursts is provided in Table~\ref{tab:bursts} with burst peak rate in the 0.4--12 keV~range, pre-burst persistent rate, the burst time scale, burst peak flux, burst fluence based on spectroscopy (refer to Section~\ref{sec:burst_spec}), and characteristic time scale that is defined as a ratio between bolometric fluence and peak bolometric flux based on spectroscopy. 
For the spectral analysis, we generated the background using the {\tt nibackgen3C50}\footnote{\url{https://heasarc.gsfc.nasa.gov/docs/nicer/tools/nicer_bkg_est_tools.html}} tool \citep{Remillard2022AJ....163..130R}. The spectral response matrix and ancillary response files were produced using {\tt nicerrmf} and {\tt nicerarf} commands, respectively. The \nicer light curve products were produced from event files using {\tt XSELECT} package.

\section{Timing studies} \label{timing-sec}

\subsection{Burst light curves} \label{sec:lc} 
We analyzed the X-ray burst light curves from \nicer in the 0.4--12~keV range at a bin time of 62.5~ms (Figure~\ref{burst_lc}). The observed bursts have peak count rates ranging from 15900 to 28200 c~s$^{-1}$ (Table~\ref{tab:bursts}). Additionally, we included the 3--12~keV burst profiles in grey in the figure. The 3--12~keV light curves show a drop around the burst peak, indicating that these bursts are of PRE nature \citep{Keek2018ApJ...856L..37K}. 
In Table~\ref{tab:bursts}, the persistent emission-subtracted burst peak rate is given along with the pre-burst (persistent) count rate. The onset detection time of each burst is reported in MJD (UTC) based on \nicer’s Mission Elapsed Time in the TT timeframe without barycentric correction. We estimated the burst duration using an ad-hoc criterion, corresponding to a time interval with a count rate greater than 1.5 times their pre-burst persistent rate (Table~\ref{tab:bursts}). 
Alternatively to the above criteria, we also determined the burst duration as an interval having a count rate of more than 10\% of the peak count rate. This resulted in similar burst durations, mostly below 10~s. Irrespective of approaches, all these bursts show a fast rise time of less than 1~s, followed by rapid decay of 4--5 s after the peak (Figure~\ref{burst_lc}). In the sample, Bursts \#1 \& \#2, and \#3 \& \#4 are the closest pair of bursts that occurred at a recurrence time of 2.17 and 2.79 hours, respectively.    
The prime composition of the accreted material can be speculated based on burst fluence and recurrence time that can be used to estimate the accretion fluence. The $\alpha$ parameter, which is defined as a ratio of accretion and burst fluences, turns out to be 187 and 235 for the above pairs of bursts, respectively. A smaller $\alpha$ of approximately 40 corresponds to higher nuclear efficiency, indicating a hydrogen-rich fuel. The parameter $\alpha$ $>$100 suggests the burning of helium-rich material, as expected in the case of \source \citep{Galloway2008ApJS..179..360G}.  An estimation of the ignition depth based on recurrence time is presented by \citet{Yu2023arXiv231216420Y}. The authors found the ignition depth in a range of (1.56--2.47)$\times$10$^8$~g~cm$^{-2}$. Moreover, following the relation between burst fluence and persistent flux, most \nicer bursts originated through the ignition of helium fuel mixed with hydrogen and heavy elements (see \citealt{Yu2023arXiv231216420Y} for details).

\begin{figure}[]
\centering
\includegraphics[height=3.1in, width=3.3in, angle=0]{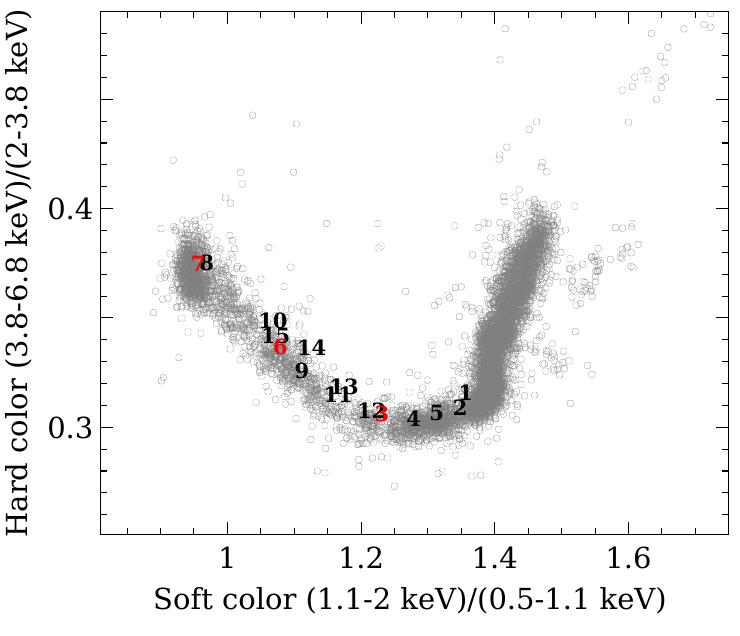}
\caption{The color-color diagram using \nicer observations of \source. The persistent emission before each burst is marked with a number in occurrence order. Two colors (black and red) are considered to distinguish overlapping burst numbers.  } 
\label{ccd}
\end{figure}

\begin{figure*}
    \centering
    \includegraphics[scale=0.4]{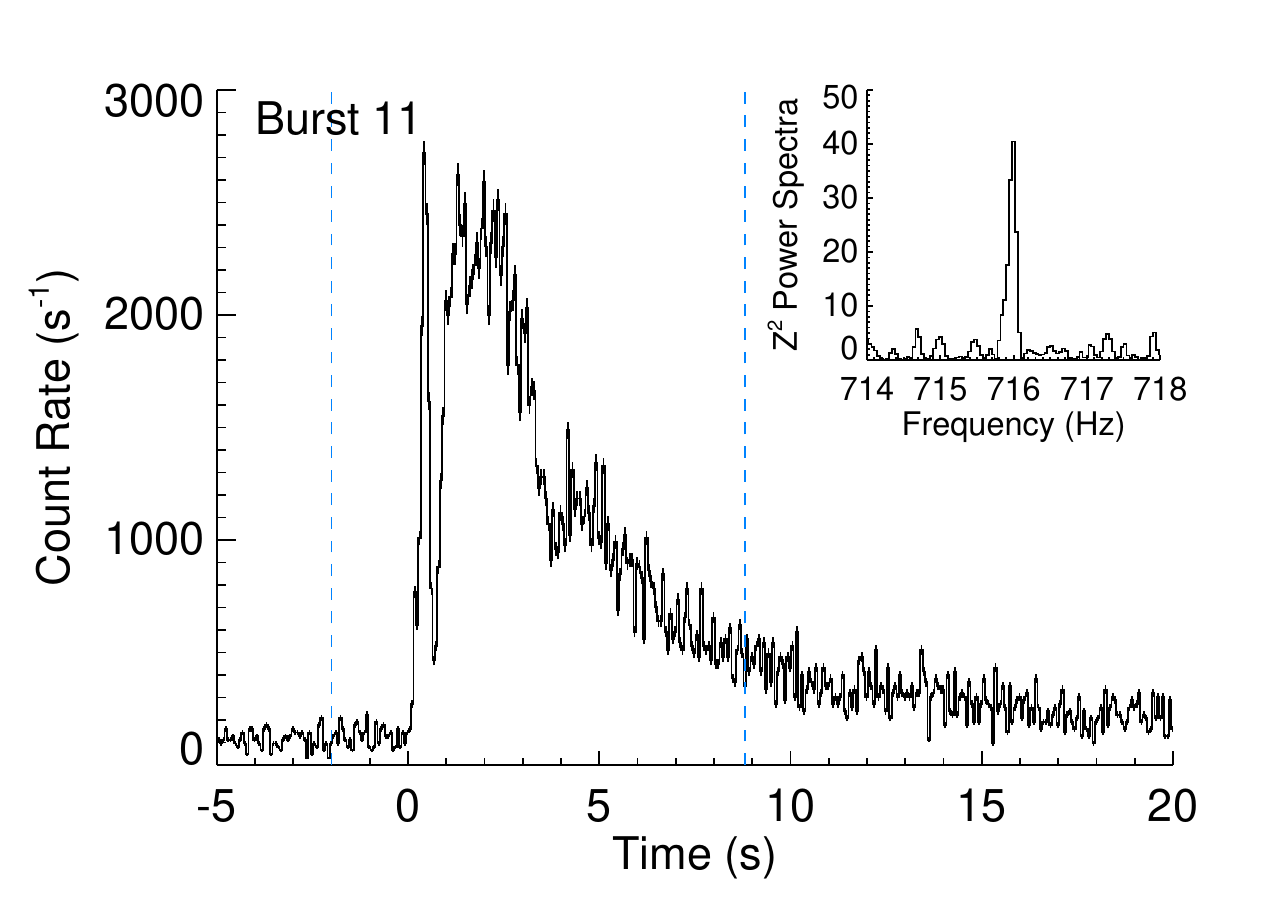}
    \includegraphics[scale=0.4]{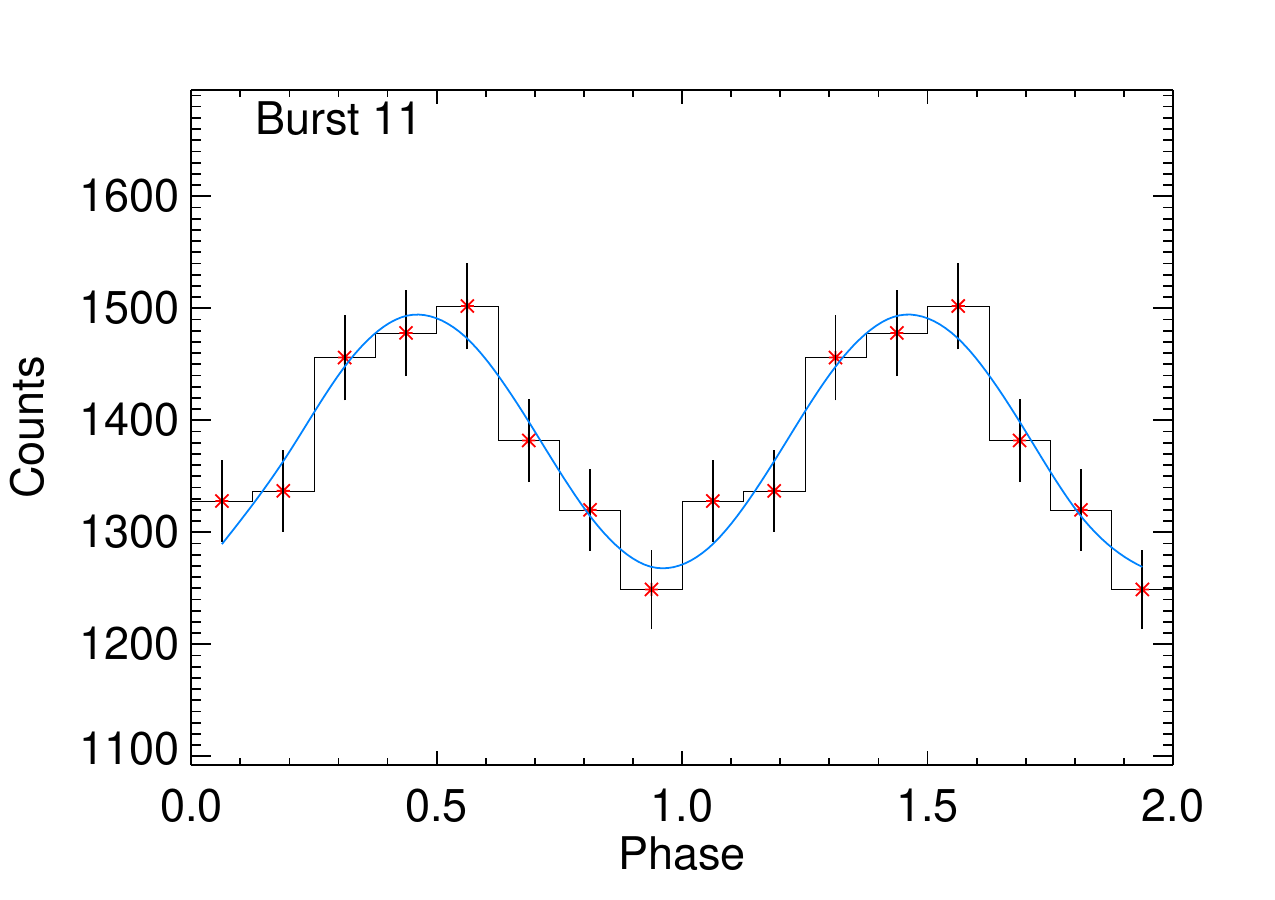} 
    \caption{Left panel shows the light curve of Burst~\#11 from \source with a bin size of 62.5~ms in 3--10~keV. Vertical blue dashed lines show the time interval used for the calculation of the $Z^2$ power spectrum for the same energy range. The inset plot refers to the power spectra in the vicinity of 716 Hz.  
    On the right, the pulse profile generated by folding the lightcurve with the 716~Hz is shown, together with the best-fitting sinusoidal model (blue line). } 
    \label{fig:b11_1dpwr}
\end{figure*}

\subsection{Color-color diagram} \label{sec:ccd} 

To study the source evolution and the spectral state before the observed bursts, we created a color-color diagram using all \nicer data accumulated between June 2017 and May 2022. We defined the soft color as the ratio of count rates in the (1.1--2.0)/(0.5--1.1)~keV energy bands, while the hard color is considered as the ratio of count rates in the (3.8--6.8)/(2.0--3.8)~keV energy bands  (see, e.g. \citealt{Bult2018, Jaisawal2019ApJ...883...61J, 2021ApJ...910...37G}). The colors before all 15 bursts are included in Figure~\ref{ccd}. 
The color-color diagram of \source in \nicer soft energy ranges is more anti-clockwise in orientation compared to the same observed using hard X-rays energy ranges above 2~keV with \rxte in previous studies (\citealt{Zand2012A&A...547A..47I, 2013MNRAS.429.3266G, Suleimanov2017MNRAS.472.3905S}). The difference in morphology is expected due to different energy bandpasses and choice of colors between \nicer and \rxte.
The \nicer bursts are located in the left branch of the diagram shown in Figure~\ref{ccd}.  This branch can resemble the island state by comparing it with the \rxte color-color diagram  \citep{Zand2012A&A...547A..47I}. Moreover, Bursts~\#7 and \#8 are located in the upper part of the island state, where \rxte detected four super-expansion bursts in an exceptional 51 days-long low-hard state of the source in 2009 \citep{Zand2012A&A...547A..47I}. 
As observed with \nicer, \source exhibits X-ray bursts when the soft color in the color-color diagram is less than 1.4  (Figure~\ref{ccd}). This bursting activity corresponds to the low state, with a persistent rate of $\leq$1800~c~s$^{-1}$ in the 0.5--10~keV range, as shown in the hardness-intensity diagram by \citet{Yu2023arXiv231216420Y}.

\subsection{Thermonuclear Burst Oscillations} \label{sec:bo}
Since there is no definite burst oscillation signal reported in the literature of \source, we performed a timing analysis based on $Zn^2$
statistics method \citep{Buccheri1983A&A...128..245B} on each burst within a 10~s time interval starting at --2~s since the burst onset. The number of harmonics is taken as n=1 for burst oscillations.  The search was conducted in a 50--1000 Hz range in three different energy ranges, namely 0.5--10, 0.5--3, and 3--10~keV.

A candidate signal was found at 716~Hz in the power spectrum of Burst~\#11, in the 3--10 keV band. We readjusted the time interval to maximize the power of the signal (\autoref{fig:b11_1dpwr}). The maximum $Z^2$ power was obtained to be 40.98 with a single trial probability of 1.26$\times$10$^{-9}$, which was calculated by assuming a Poisson noise distributed as $\chi^2$ with two degrees of freedom.  We estimated the chance occurrence probability of this signal to be $5.4\times 10^{-4}$. This is obtained by multiplying the single trial probability and a total number of search trials 4.275$\times$10$^{5}$ performed in the 50--1000~Hz frequency range at a resolution of 0.1~Hz. The chance occurrence probability corresponds to the signal significance of 3.46$\sigma$.

We next tested this finding using other methods, such as the Leahy normalized Fast Fourier and Lomb-Scargle periodograms. We were able to reproduce similar results with a candidate oscillation at 716~Hz from Burst~\#11 using all three methods. Note that we found a Leahy power of 37.6 at 716~Hz for Burst~\#11, in the 3--10 keV band while performing a search in the 100-1000~Hz range at a frequency resolution of 0.1~Hz. 

We also computed the fractional rms amplitude of the detected signal by fitting a sinusoidal model ($A + B \sin(2 \pi \nu t - \phi_0)$) to the phase folded light curve, which is calculated in the time window where the signal detected. We obtained the fractional rms amplitude from the best fitting parameters, defined as $B/(\sqrt{2}A)$. This way, the rms amplitude of the signal during Burst~\#11 is found as 5.8\%$\pm$0.7\% in the 3--10~keV band (\autoref{fig:b11_1dpwr}). It is within the characteristic range observed in other burst oscillation sources \citep{Bhattacharyya2022ASSL..465..125B}.

\begin{figure}
    \centering  
    \includegraphics[width=\columnwidth]{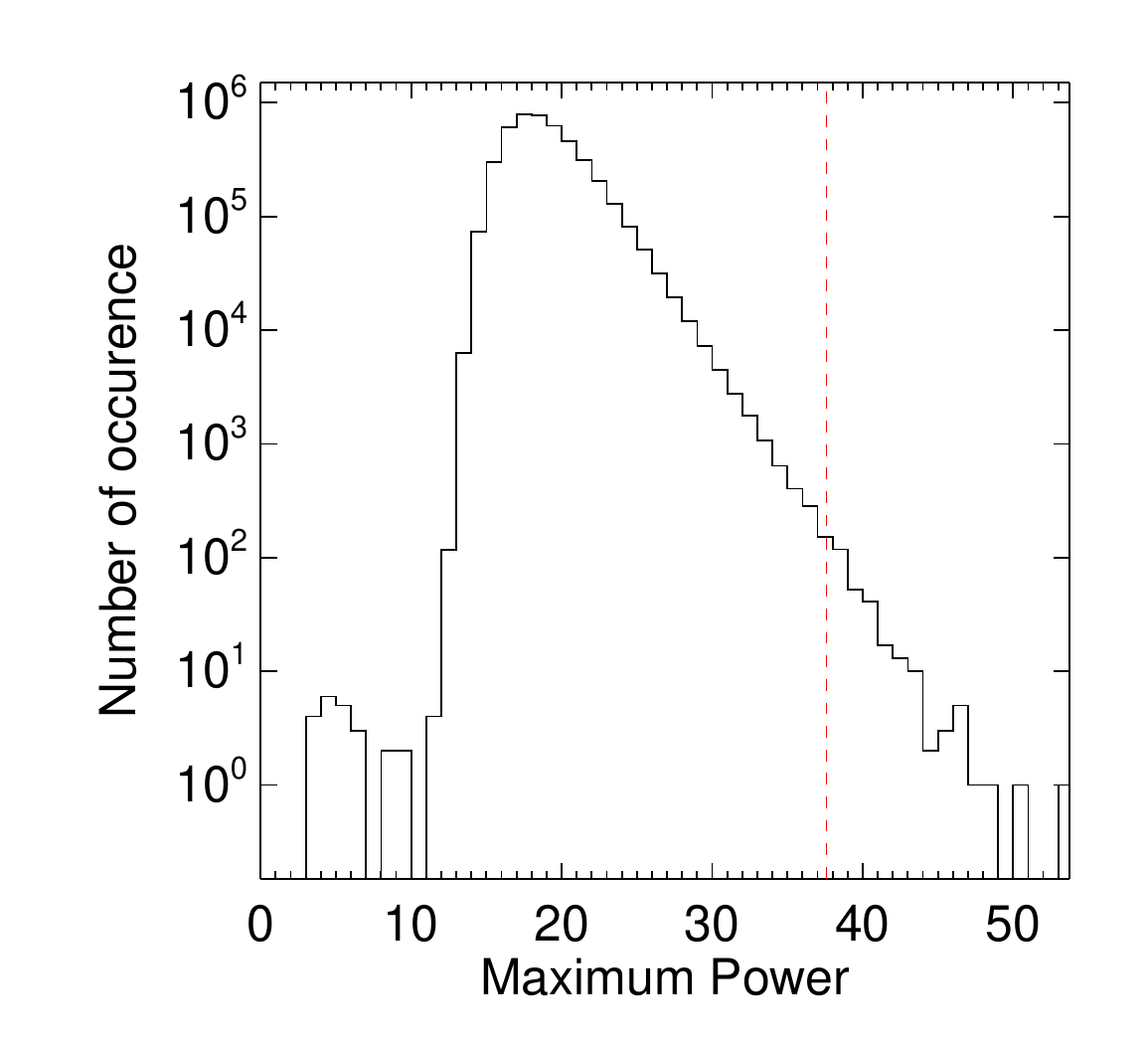}
    \caption{Distribution of maximum powers obtained from simulations. The Red dashed line shows the Fourier power value of 37.6 of the signal at 716 Hz found from Burst~\#11 in the 3--10 keV range.}
    \label{fig:dist_sim}
\end{figure}

To ascertain the significance of this candidate oscillation, we generated $10^5$ Monte Carlo simulations for each burst and each energy band. Following the recipe prescribed in \cite{Bostanci2023ApJ...958...55B}, simulations were produced by generating random times of arrival that matched the total count rate (accounting for its uncertainty, as well) observed within ~0.1 s windows during each burst. This approach ensures that our simulations closely resemble the real data. Given the significant time required by the $Z^2$ method (approximately five minutes per simulation), we opted for the Fast Fourier Transformation approach in our analysis and restricted the simulations only in the 100--1000~Hz as opposed to 50--1000~Hz, to avoid the effects of red noise on the simulated power spectra. We calculated the Leahy-normalized power spectrum for each simulated light curve with a binning of 1/2048~s and determined the maximum power. The number of occurrences of power values above the maximum Leahy-normalized power value of 37.6 (corresponding to that of Burst~\#11) for each burst and energy interval is given in \autoref{simre_table}. In the case of Burst~\#11 in 3--10 keV, there were only 12 such instances out of $10^5$ simulations (\autoref{simre_table}). The distribution of maximum powers obtained from simulations of all 15 bursts is shown in Figure~\ref{fig:dist_sim}. 

We calculated the significance of the search for all bursts in the three energy bands by dividing the total number of occurrences, i.e., 395, by $10^5$, resulting in a value of $3.95\times 10^{-3}$ from the simulation. This corresponds to a significance of 2.88$\sigma$, which is slightly lower than the chance occurrence probability of $2.77\times 10^{-3}$ or a significance of 2.99$\sigma$ calculated from the maximum Leahy power of the signal. The chance occurrence probability from Leahy power of 37.6 was estimated by multiplying the value of single trial probability and a total number of search trials 4.05$\times$10$^{5}$ performed in the 100--1000~Hz frequency range at a resolution of 0.1~Hz. From Table~2, we observed that the simulation results in a higher value of chance occurrence in the 3-10 keV energy band compared to the other two energy bands. This is likely due to the strong variability in the bursts, which skews the noise power distribution, especially in the hard band (see also \citealt{2001MNRAS.321..776F}). Furthermore, the \nicer effective area significantly decreases above 3 keV, which can also contribute to the observed uncertainties in the higher energy range. These effects can explain the lower significance obtained from the simulation.


\begin{figure}
    \centering
    \includegraphics[scale=0.4]{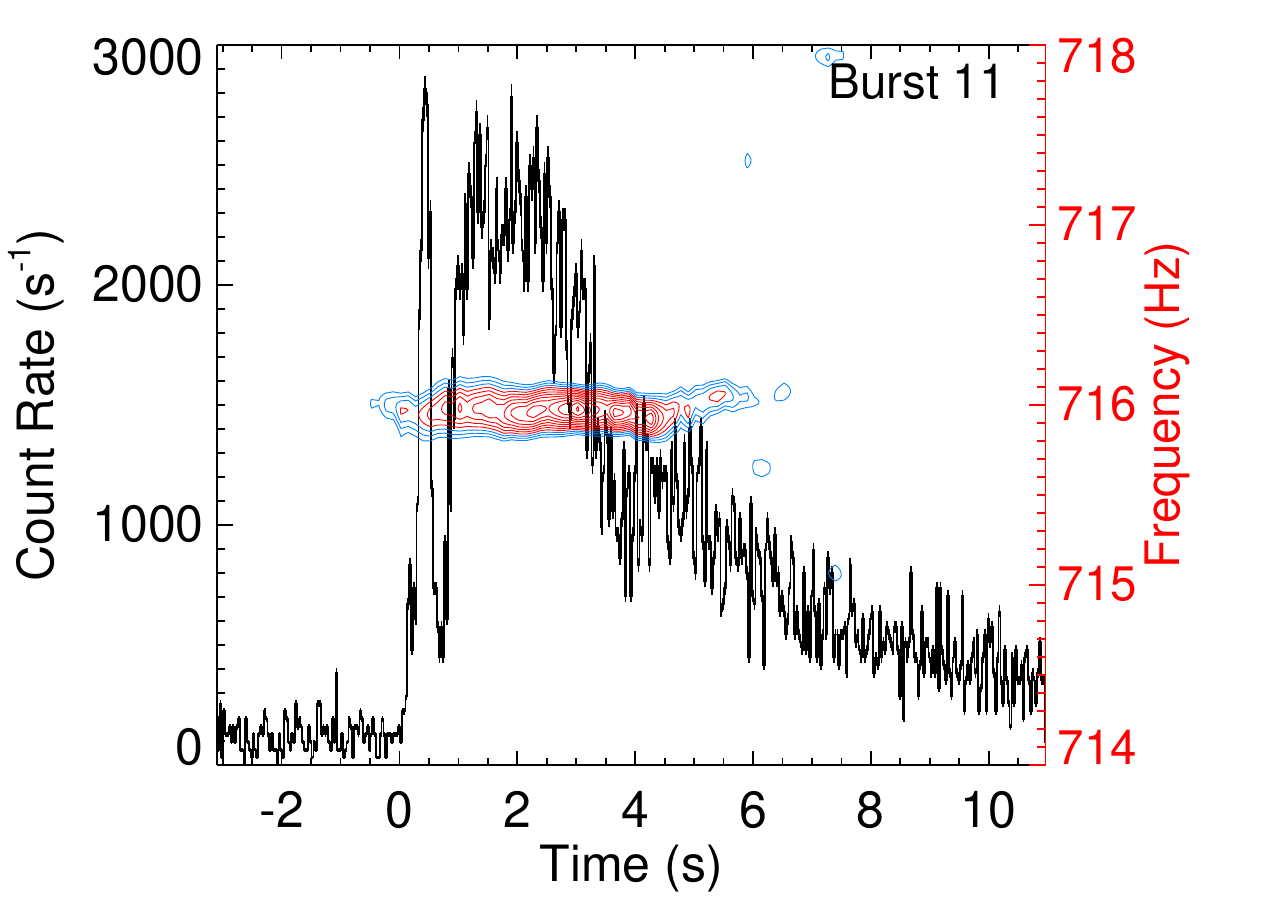}
\caption{The 3--10 keV X-ray burst light curve of Burst~\#11 at a bin size of 0.125 s (black). The contours of dynamical power spectra showing the burst oscillations at 716~Hz are included. Contours refer to $Z^{2}>10$ (blue) and $>15$ (red) up to the maximum with steps of 2. 
}
\label{fig:b11_osc}
\end{figure}
 
\begin{table}[]
    \centering
    \caption{Number of simulations for each burst and energy interval where a signal showing a power equal to or larger than the Fourier power with Leahy-normalized value of 37.6 out of 10$^5$ simulations.}
    \begin{tabular}{cccc}
    \hline
       Burst  & \multicolumn{3}{c}{Energy Range} \\
       Number & 0.5--10~keV  & 0.5--3~keV & 3--10~keV \\
       \hline
        1 &10 & 6& 11\\
        2 &4 & 9& 12\\
        3 & 7 & 8 & 7 \\
        4 & 5 & 9 & 18 \\
        5 & 3 & 10 & 14 \\
        6 & 8 & 7 & 17 \\
        7 & 9 & 7 & 11 \\
        8 & 7 & 8 & 14 \\
        9 & 3 & 4 & 13 \\
        10 & 6 & 8 & 11 \\
        11 & 8 & 12 & 12 \\
        12 & 5 & 5 & 12 \\
        13 & 7 & 7 & 10 \\
        14 & 6 & 6 & 17 \\
        15 & 3 & 6 & 13 \\
        \hline
    \end{tabular}
    \label{simre_table}
\end{table}

We also checked for peaks at 280~Hz \citep{Zand2012A&A...547A..47I, Bilous2019} and 358~Hz (half of 716 Hz) and did not find any significant powers around these frequencies. We next constructed a dynamic power spectrum through Burst~\#11 using the $Z^2$ method. Here, we considered time intervals between --5 and 10 seconds with respect to the burst onset and frequencies between 714--718~Hz with a step of 0.01~Hz in the energy band of 3--10 keV, and used a time window of 4~s, shifted with a time step of 1/8~s. The resulting $Z^2$ contour map with the light curves is given in Figure~\ref{fig:b11_osc}. Burst~\#11 shows an oscillatory signal between the rising and decaying parts of the burst with no apparent shift. 

We attempted another independent approach to examine the 716~Hz candidate signal in Burst~\#11. For this, the burst was divided into two non-overlapping time bins in the 3--10~keV range, each with a total of 5000 counts by following the procedure described by \citet{Ootes2017ApJ...834...21O} and \citet{Galloway2020ApJS..249...32G}. The first and second bins correspond to a time range between 0--2.8~s and 2.8--7.5~s of the burst light curve as shown in the left panel of \autoref{fig:b11_1dpwr}, respectively. We searched for the candidate signal in a narrow frequency range between 711--721~Hz ($\nu_0\pm$5~Hz) at a frequency resolution of 1~Hz using the $Z^2$ statistics method \citep{Ootes2017ApJ...834...21O}.  The maximum $Z^2$ power was found to be 18.01 and 17.31 in the first and second intervals, with single trial probabilities of $1.23\times 10^{-4}$ and $1.74\times 10^{-4}$, respectively. Our finding matches the criterion~3 based on the double bin detection as defined in \citet{Ootes2017ApJ...834...21O}. We used their equations~3 and 4 to calculate the noise chance probabilities of the two individual adjacent bins. The chance probabilities are estimated to be $8.1\times 10^{-6}$ and $2.45\times 10^{-6}$, suggesting independent detections of the 716~Hz signal in Burst~\#11.

Lastly, we remark on the null results on burst oscillations reported by \citet{Yu2023arXiv231216420Y} using the same \nicer data. The authors have used only a fixed 4~s window moving with 0.5~s steps by applying the Fast Fourier transform with Leahy Normalization. They considered three narrow energy ranges such as 0.3--3, 3--6, and 6--12~keV for search without taking into account the change in the effective area of \nicer. The exact time interval for the oscillation search, including the search on pre-burst emission, is also not discussed in their study. Considering the above factors, it is likely that \citet{Yu2023arXiv231216420Y} did not find any burst oscillations from \source.

\begin{table*}[ht!]
\centering
\caption{Spectral Parameters of the observed persistent emission before each burst from \source. The pre-burst emission is fitted with an absorbed Comptonization model with a blackbody component.}
\begin{tabular}{cccccccccc}
\hline 
Burst& N${\rm_H}$ &kT$\rm_{inner}$ &Norm$\rm_{inner}$ &Radius$\rm_{inner}$  &kT$\rm_{comp}$ & kT$\rm_{e}$ &$\tau$ &Flux$^*$     &$\chi^2$/dof\\
Number & (10$^{21}~\rm{cm}^{-2}$) &(keV) &    &(km) &(keV) &(keV)  \\
\hline
1	&2.4$\pm$0.1    &0.51$\pm$0.03  &851$\pm$279      &24.5$\pm$4	&0.05$\pm$0.02  &2.6$\pm$0.4	&7.5$\pm$0.7	 &6.3$\pm$0.1   &649/605\\
2	&2.5$\pm$0.1    &0.50$\pm$0.01  &882$\pm$105      &24.5$\pm$1.5	&0.05$\pm$0.02  &2.9$\pm$0.3	&7.0$\pm$0.9	 &6.2$\pm$0.1   &937/827\\
3	&2.5$\pm$0.1    &0.42$\pm$0.02  &1243$\pm$196     &29.6$\pm$2.3	&0.05$\pm$0.02  &3.3$\pm$1.2	&5.9$\pm$1	 &4.9$\pm$0.1       &933/846\\
4	&2.4$\pm$0.1    &0.44$\pm$0.01  &1307$\pm$222     &30.4$\pm$2.6	&0.04$\pm$0.01  &3.6$\pm$1.1	&6$\pm$0.9	 &5.3$\pm$0.1       &723/701\\
5	&2.5$\pm$0.1    &0.47$\pm$0.01  &1073$\pm$115     &27.5$\pm$1.5	&0.05$\pm$0.01  &3.1$\pm$0.2	&6.6$\pm$0.3	 &5.7$\pm$0.1   &964/835\\
6	&2.1$\pm$0.1    &0.30$\pm$0.01  &4167$\pm$272     &54.2$\pm$1.8	&0.09$\pm$0.02  &3.5$\pm$0.4	&6.3$\pm$0.4	 &3.2$\pm$0.1   &950/810\\
7	&2.1$\pm$0.1    &0.17$\pm$0.01  &20540$\pm$8731   &120.4$\pm$25.6	&0.06$\pm$0.02  &3.1$\pm$0.5	&6.9$\pm$0.5	 &2.1$\pm$0.1  &672/614\\
8	&2.0$\pm$0.1    &0.19$\pm$0.01  &11914$\pm$3373   &91.7$\pm$13	&0.07$\pm$0.01  &3.0$\pm$0.3	&7.1$\pm$0.4	 &2.2$\pm$0.1   &835/703\\
9	&2.0$\pm$0.1    &0.33$\pm$0.01  &3104$\pm$282     &46.8$\pm$2.1	&0.1$\pm$0.01  &5.3$\pm$1	&5.2$\pm$1	 &3.3$\pm$0.1           &788/722\\
10	&2.0$\pm$0.1    &0.29$\pm$0.01  &4622$\pm$461	  &57.1$\pm$2.8  &0.08$\pm$0.01  &4.2$\pm$1	&5.8$\pm$0.7	 &3.1$\pm$0.1       &881/768\\
11	&2.2$\pm$0.1    &0.34$\pm$0.01  &2832$\pm$256     &44.7$\pm$2	&0.03$\pm$0.01  &5.4$\pm$1.4	&4.8$\pm$1.5	 &3.7$\pm$0.1   &653/662\\
12	&2.3$\pm$0.1    &0.38$\pm$0.01  &1949$\pm$100     &37.1$\pm$1 	&0.04$\pm$0.01  &4.2$\pm$0.7	&5.5$\pm$0.6	 &4.1$\pm$0.1   &741/746\\
13	&2.0$\pm$0.1    &0.35$\pm$0.01  &3088$\pm$243	  &46.7$\pm$1.8 &0.09$\pm$0.01  &5.6$\pm$1	&5.1$\pm$1	 &3.8$\pm$0.1           &780/733\\
14	&1.7$\pm$0.2    &0.34$\pm$0.01  &3530$\pm$514	  &49.9$\pm$3.6    &0.12$\pm$0.01  &6.3$\pm$1.5	&5.3$\pm$0.8	 &3.6$\pm$0.1   &673/662\\
15	&1.9$\pm$0.1    &0.29$\pm$0.01  &4451$\pm$565     &56$\pm$3.6	&0.09$\pm$0.01  &3.5$\pm$0.9	&6.5$\pm$0.6	 &3.1$\pm$0.1   &745/686\\
\hline
\end{tabular}
\label{tab:pre-burst}\\
\footnotesize{Note: The pre-burst blackbody temperature, normalization, and radius (in km, calculated at a distance of 8.4 kpc; \citealt{Valenti2004MNRAS.351.1204V, Keek2018ApJ...856L..37K}) from this blackbody component are denoted with kT${\rm _{inner}}$, Norm${\rm_{inner}}$, and Radius${\rm_{inner}}$, respectively. \\ 
From Comptonization model, kT$\rm_{comp}$: input soft photon temperature, kT$\rm_{e}$: electron plasma temperature, $\tau$: plasma optical depth.\\  
$^*$ 0.5--10 keV unabsorbed flux in units of 10$^{-9}$~erg~s$^{-1}$~cm$^{-2}$. }\\
\end{table*}

\begin{figure*} 
\centering
	{\includegraphics[scale=0.54]{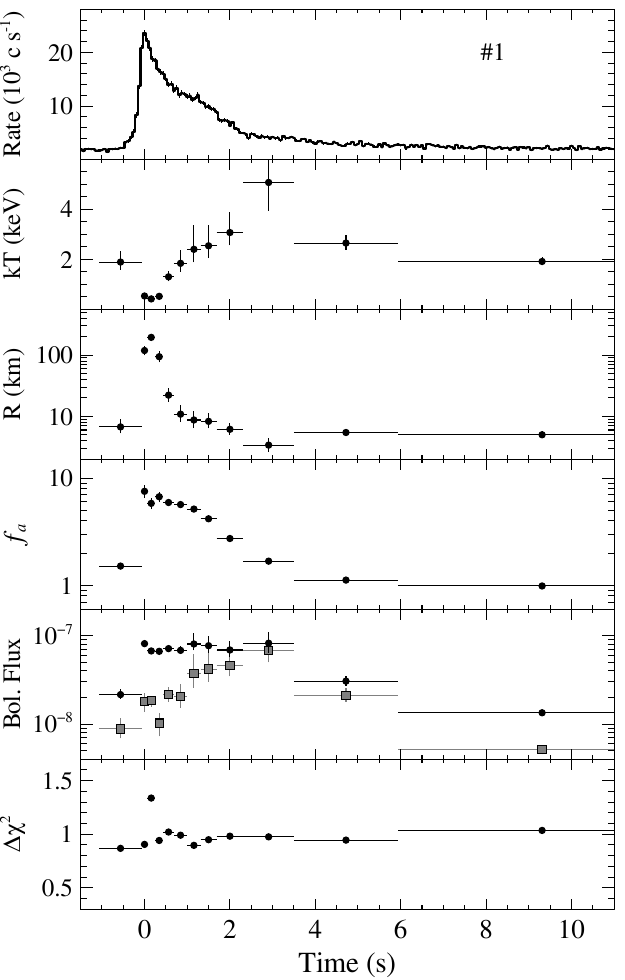}}
	{\includegraphics[scale=0.54]{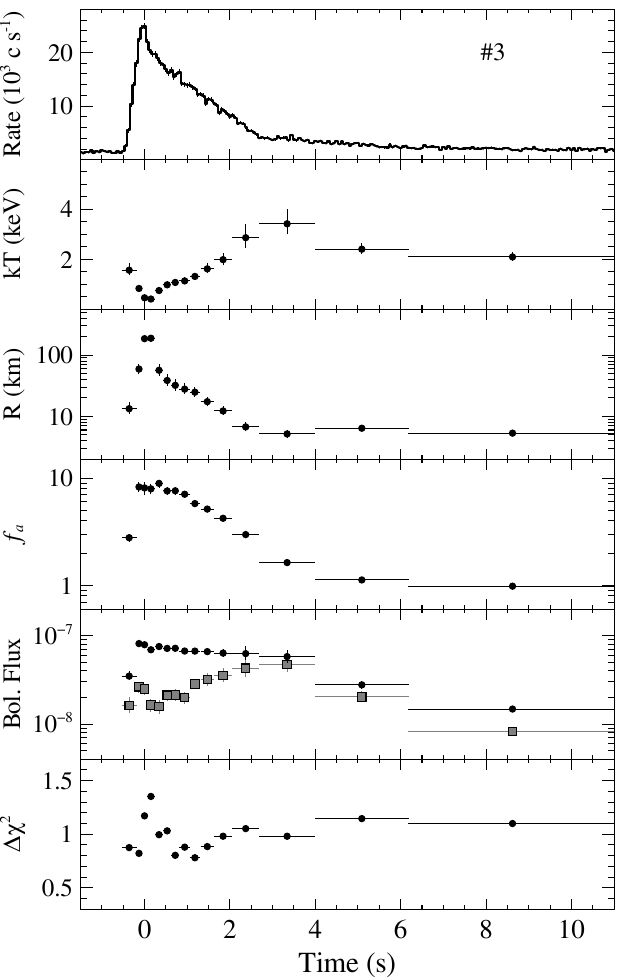}}
		{\includegraphics[scale=0.54]{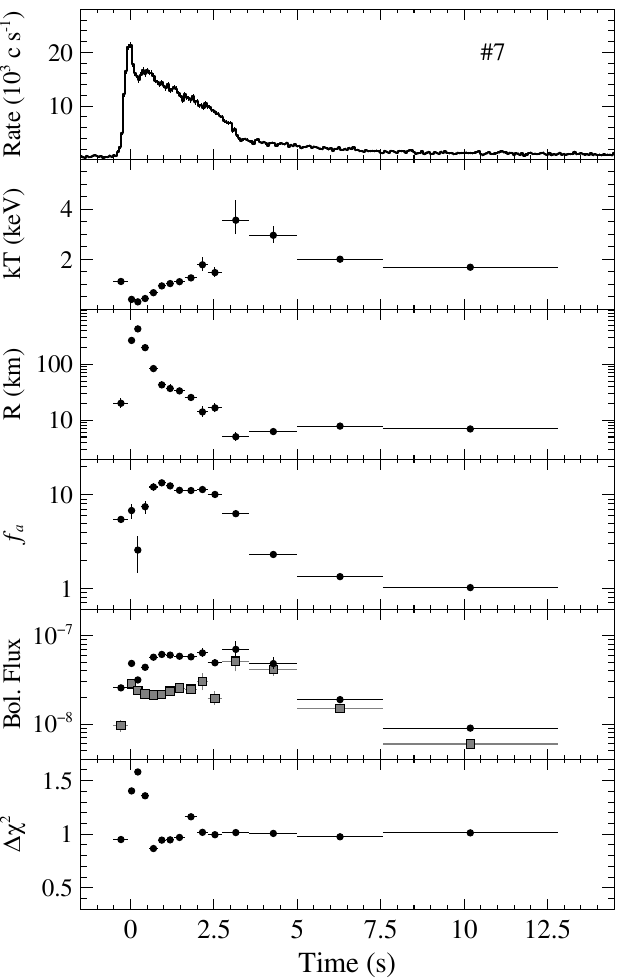}} \\ 
		{\includegraphics[scale=0.54]{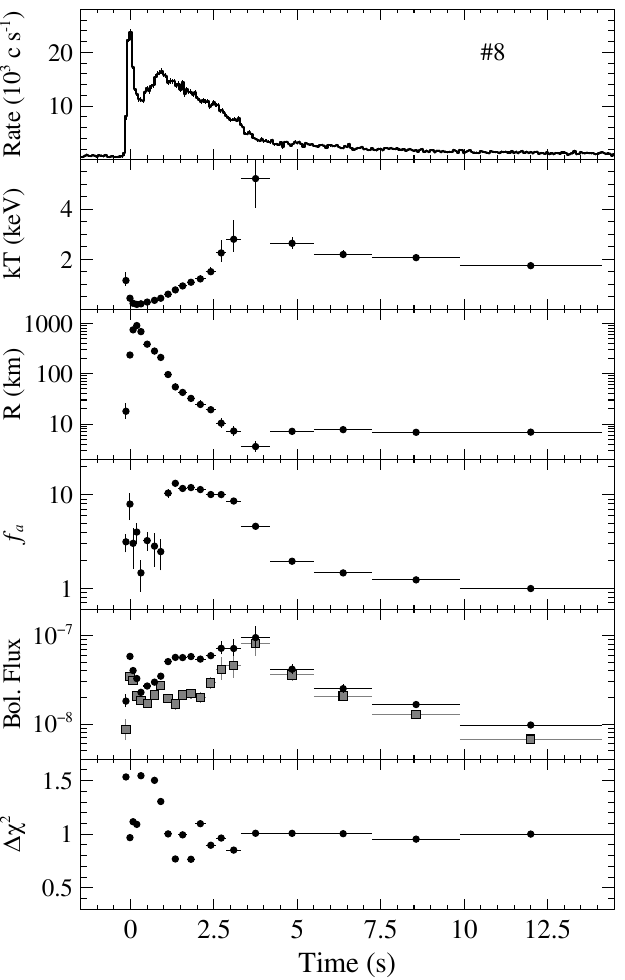}}
		{\includegraphics[scale=0.54]{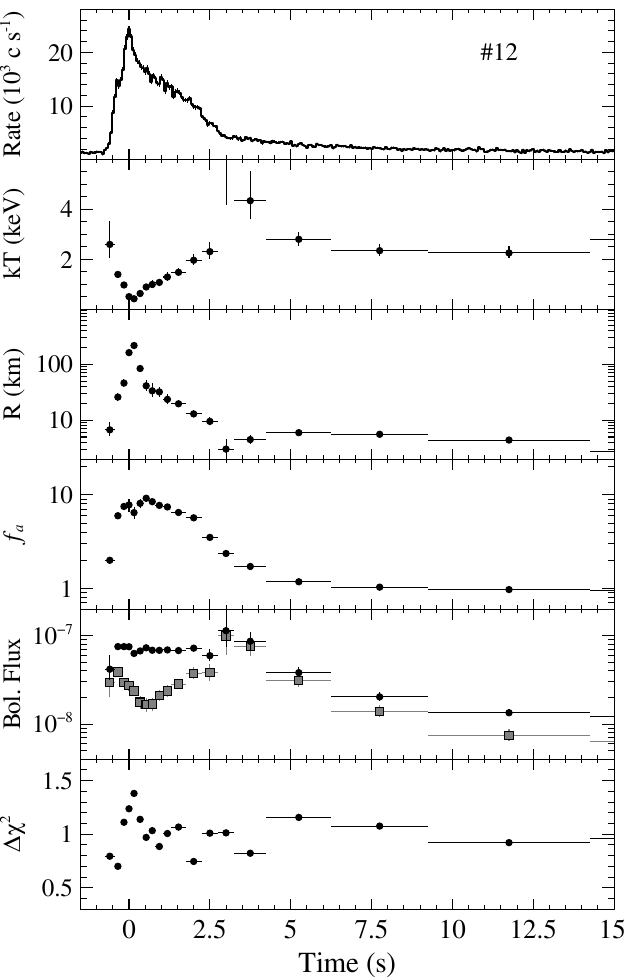}}
		{\includegraphics[scale=0.54]{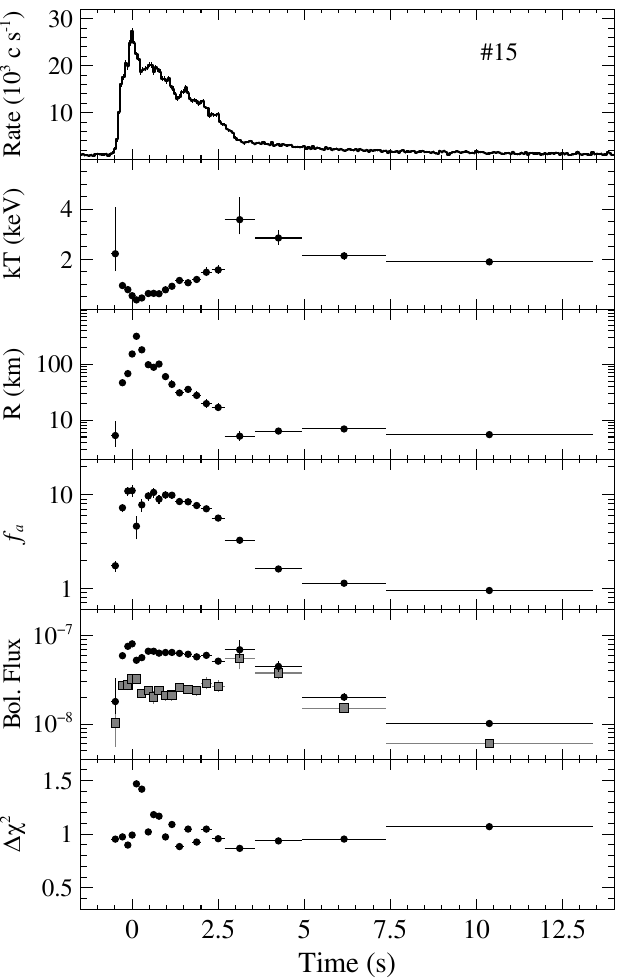}} 
    \caption{Spectral parameters obtained from time-resolved spectroscopy of X-ray bursts from \source using a variable persistent emission model. The evolution of the six most representative bursts is shown. The black circle and grey square in the fifth panel indicate the total and the blackbody bolometric fluxes in the 0.1--100~keV range, respectively. }
    \label{fig:burst_spec1}
\end{figure*}

\section{Spectroscopy} \label{spec-sec}

Our findings on pre-burst accretion emission and burst time-resolved spectroscopy based on a variable persistent emission method, an alternative method (a variable pre-burst component method),  and a physical reflection model are presented in this section. 

\subsection{Pre-burst persistent emission of \source} \label{sec:per-spec}
The persistent emission before each burst is studied before performing burst time-resolved spectroscopy.  The exposure time of the pre-burst emissions ranged between 125 and 1000~s depending on data available in the same \nicer orbit prior to the burst. 
Following  \citet{Keek2018ApJ...856L..37K}, we fitted the persistent emission using an absorbed Comptonization model ({\tt Comptt} in {\tt XSPEC}; \citealt{Titarchuk1994ApJ...434..570T}) with a blackbody component in our study. The absorption column density is modeled with a {\tt Tbabs} component in {\tt XSPEC} with {\tt wilm} abundance \citep{Wilms2000} and {\tt verner} cross-section \citep{Verner1996ApJ...465..487V}.  The best-fitted spectral parameters are given in Table~\ref{tab:pre-burst}. The 0.5--10 keV unabsorbed persistent flux was found to be in a range of (2.1---6.3)$\times$10$^{-9}$~erg~s$^{-1}$~cm$^{-2}$ before the bursts. The source luminosity can be calculated to be in a range of (1.77---5.33)$\times$10$^{37}$~erg~s$^{-1}$ at a distance of 8.4~kpc,   suggesting the persistent emission to be 5 to 16\% of the Eddington limit of a pure helium accreting NS.  

The absorption column density ($\rm N_H$) along the line of sight of the source was detected at an average value of 2.2$\times$10$^{21}$~cm$^{-2}$, in good agreement with earlier findings \citep{2010ApJ...719.1807G}. We noticed the blackbody temperature from pre-burst emission varies in a range of 0.17 to 0.51 keV. Moreover, the corresponding emission radii remain larger than the neutron star radius, which may represent the emission from the inner part of the accretion disk.  The errors on spectral parameters are calculated for a 90\% confidence interval in this paper unless specified otherwise.  The convolution {\tt cflux} model in {\tt XSPEC} was used for flux measurement in this paper.  

In addition to the above model, a combination of an absorbed disk blackbody model and a power-law component can equally describe the pre-burst emission. Choosing another persistent model does not have any noticeable effect on the burst spectral evolution or other correlation studies such as presented in  Figure~\ref{fig:burst_para_rel}.

\subsection{Burst time-resolved spectroscopy using a variable persistent emission method } \label{sec:burst_spec}

For burst time-resolved spectroscopy, we extracted spectra at finer time intervals for all 15 observed bursts to understand their burst evolution. Each time segment was chosen so that the spectrum contained a minimum count of 3000 in the 0.4--10~keV range. 
A typical number of time-resolved intervals ranges from 12 to 20 with a median of 18 segments, each having an exposure time between 0.125 to 2~s.
We first attempted to describe the burst spectral segments with a simple blackbody component. This resulted in an unacceptable fit with a reduced chi-square of more than 2 because of excess spectral residuals with different slopes observed below and above 2 keV. Therefore, we used the variable persistent emission method (also known as $f_a$-model; \citealt{Worpel2013, Worpel2015}) to fit the burst spectra. In this approach, a fixed pre-burst emission is allowed to vary only through a multiplicative factor that incorporates any changes in the pre-burst emission during the burst. Moreover, the main burst emission is approximated by a thermal blackbody component. We found that the  variable persistent emission method can well describe time-resolved spectra of all bursts with a reduced chi-square value of $<$1.5. Figure~\ref{fig:burst_spec1} shows the evolution of spectral parameters such as burst temperature and the corresponding thermal radius (R in km, calculated assuming a distance of 8.4~kpc), the persistent multiplicative scaling factor ($f_a$), the total (black circle) and blackbody (grey square) bolometric fluxes in 0.1--100 keV, and reduced chi-square ($\Delta\chi^2$) along with 0.4--12~keV burst light curve in the top panel, for six thermonuclear X-ray bursts (for representation purposes).   

All 15 X-ray bursts showed clear signs of PRE where the blackbody temperature approached a minimum value at the burst peak, while the photospheric radius reached a maximum value around this point at a constant total flux level (see Figure~\ref{fig:burst_spec1}). A maximum blackbody temperature of about 4~keV was observed during these bursts at the touchdown when the NS photosphere returned to the surface. Except for Burst~\#2, the bursts exhibit strong expansion with a radius of more than 100~km, measured assuming a distance of 8.4~kpc. A maximum photospheric radius of 902$\pm$103~km was observed for Burst~\#8. A burst with large expansion radii of $\ge$1000~km is usually considered to be a super-expansion burst \citep{Zand2012A&A...547A..47I, Yu2023arXiv231216420Y}. This is the first time with \nicer where we observed a very large expansion of 902$\pm$103~km, indicating the detection of a super-expansion burst from \source (see, e.g. \citealt{Zand2012A&A...547A..47I} for super-expansion bursts with \rxte). The lowest temperature of 0.2~keV was found at the peak of this burst.


\begin{figure*}
\centering
		{\includegraphics[scale=0.55]{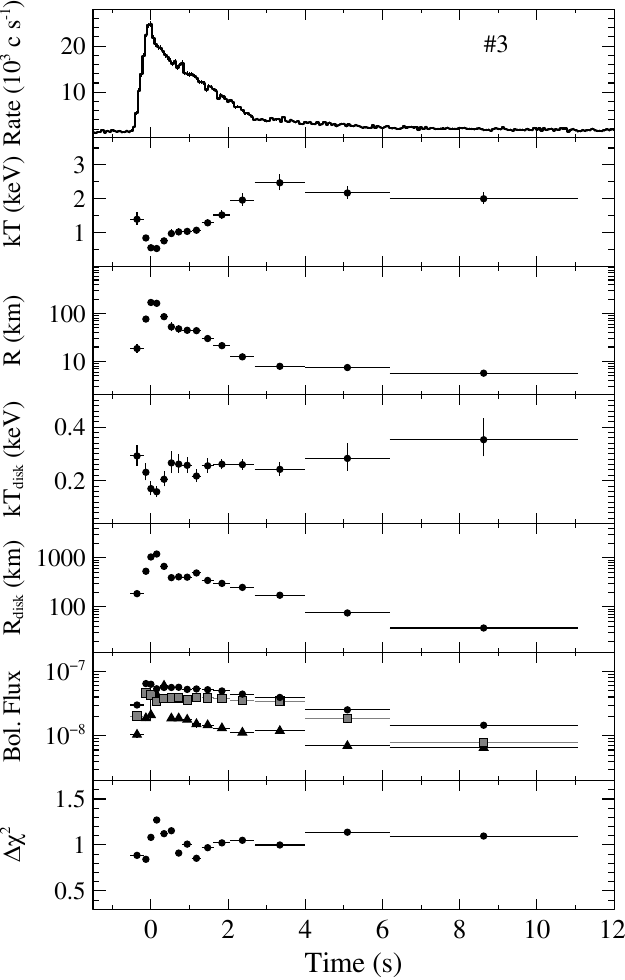}}
		{\includegraphics[scale=0.55]{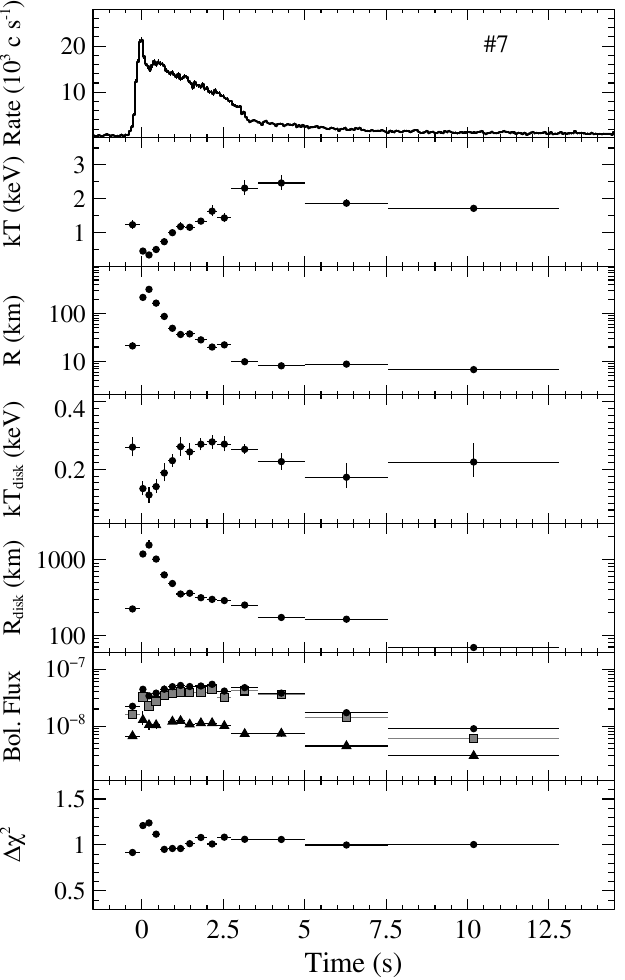}}
		{\includegraphics[scale=0.55]{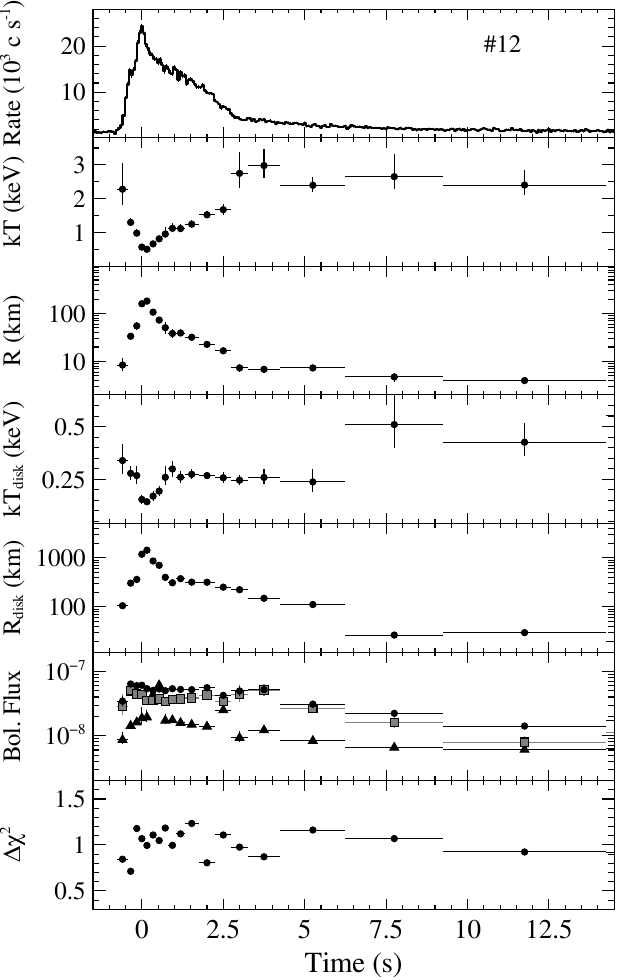}}
    \caption{Spectral parameters obtained from time-resolved spectroscopy of three representative bursts from \source using the second modeling approach. The black circle, grey square, and black triangle in the sixth panel indicate the total model flux, burst blackbody flux, and the second blackbody flux from the disk in the 0.1--100~keV range, respectively.}
    \label{fig:burst_spec_2pervar}
\end{figure*}

From Figure~\ref{fig:burst_spec1}, the scaling factor \fa evolved within the burst time scale, reaching its maximum around the peak of the burst light curve. The highest value of \fa $>$13 was observed for Bursts~\#7 and \#8, indicating the persistent emission level is enhanced by more than an order of magnitude in these bursts. 
The total unabsorbed bolometric flux represents a sum of fluxes from burst and evolving accretion emission. It peaks in a range of (7--11.4)$\times$10$^{-8}$~erg~s$^{-1}$~cm$^{-2}$ and remains plateau for a few seconds before declining after the burst touchdown (Figure~\ref{fig:burst_spec1}). On the other hand, the flux from the main blackbody component i.e. burst emission, reaches its maximum value at the touchdown and is found between (4.5--10)$\times$10$^{-8}$~erg~s$^{-1}$~cm$^{-2}$ from these bursts (Figure~\ref{fig:burst_spec1}). The value matches well with the Eddington limited flux of (6.05$\pm$2.26)$\times$10$^{-8}$~erg~s$^{-1}$~cm$^{-2}$, as per the Multi-INstrument Burst ARchive (MINBAR)  studies of 65 PRE bursts from \source \citep{Galloway2020ApJS..249...32G}.

As an alternative to the  variable persistent emission method, we attempted to investigate the changes in the persistent model parameters during these bursts. For this, we allowed one of the components of the persistent emission to vary during the burst with the fixed column density  (second approach or a variable pre-burst component method). Our pre-burst model consisted of an absorbed Comptonization model with a blackbody component. We permitted the blackbody parameters to change. This component may originate from the inner part of the accretion disk and is suitable for evaluating the burst-disk interaction.  Similar to the variable persistent emission model (Figure~\ref{fig:burst_spec1}), we observed an equivalent variation in the burst parameters as shown in Figure~\ref{fig:burst_spec_2pervar} for three representative bursts, using the second modeling approach. On the other hand, the (pre-burst) blackbody temperature (kT$_{disk}$) was observed to be relatively cooler, with an emission radius (R$_{disk}$) of more than 100 km in the rising phase of the burst. The lowest temperature value was detected at the burst peak, where the emission radius from the inner disk was found to be $\ge$1000 km, depending on the burst. The change in disk radius from 100 to 1000~km within a couple of seconds appears like a rapid expansion. A study by \citet{Fragile2020} suggests that the disk may retreat by a few tens of km and recover rapidly on the burst time scale. The response of the disk depends on the luminosity as well as the disk viscosity. In our study, the observed behavior may suggest that the inner part of the accretion disk gets affected during the bursts. Being strongly irradiated by the burst emission, the inner part of the disk is expected to retreat mostly due to Poynting-Robertson drag-driven accretion \citep{Fragile2020}. At the same time, the optical depth of the accretion disk may decrease by an order of magnitude \citep{Fragile2020}.  A lower optical depth may allow the burst photons to interact with the disk to a larger extent, hence it may appear like an expansion. However, it is still unclear how the inner disk could cool down promptly when it is supposed to be strongly heated by the burst. Therefore, to understand this, we studied the spectral evolution of these bursts using a reflection model in the next section.


\begin{figure*}
\centering
		{\includegraphics[scale=0.55]{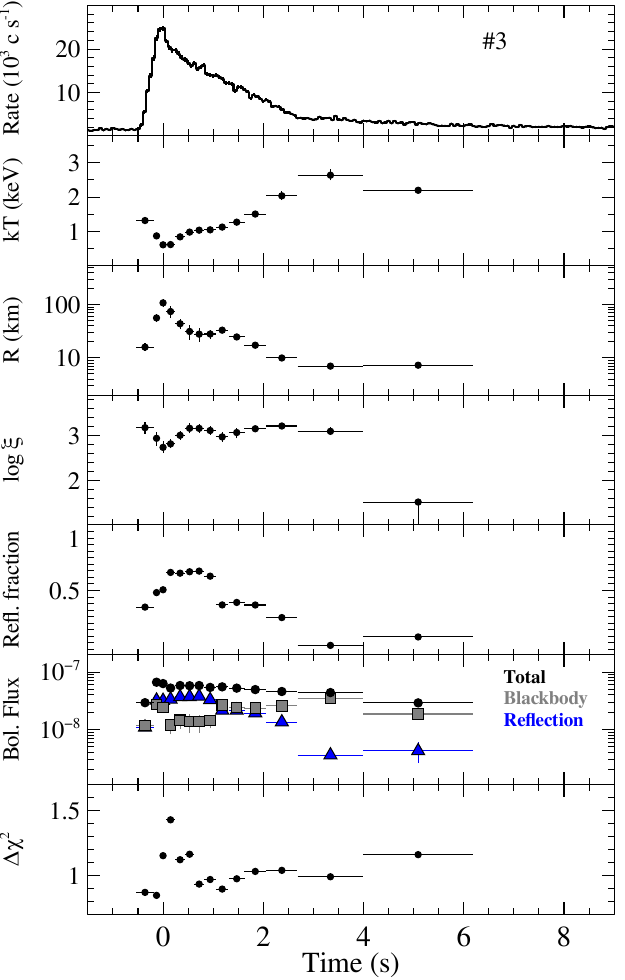}}
		{\includegraphics[scale=0.55]{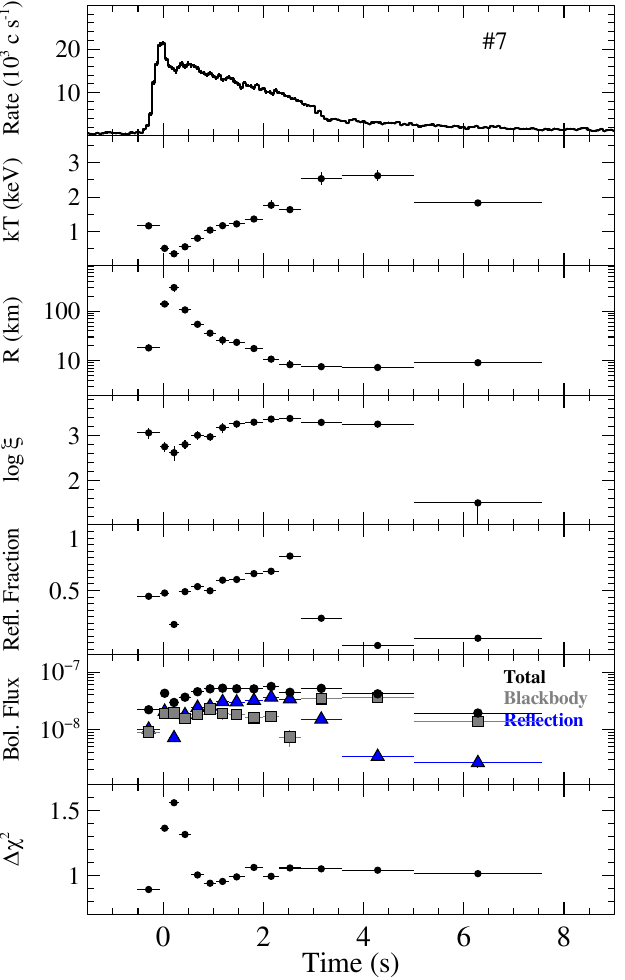}}
		{\includegraphics[scale=0.55]{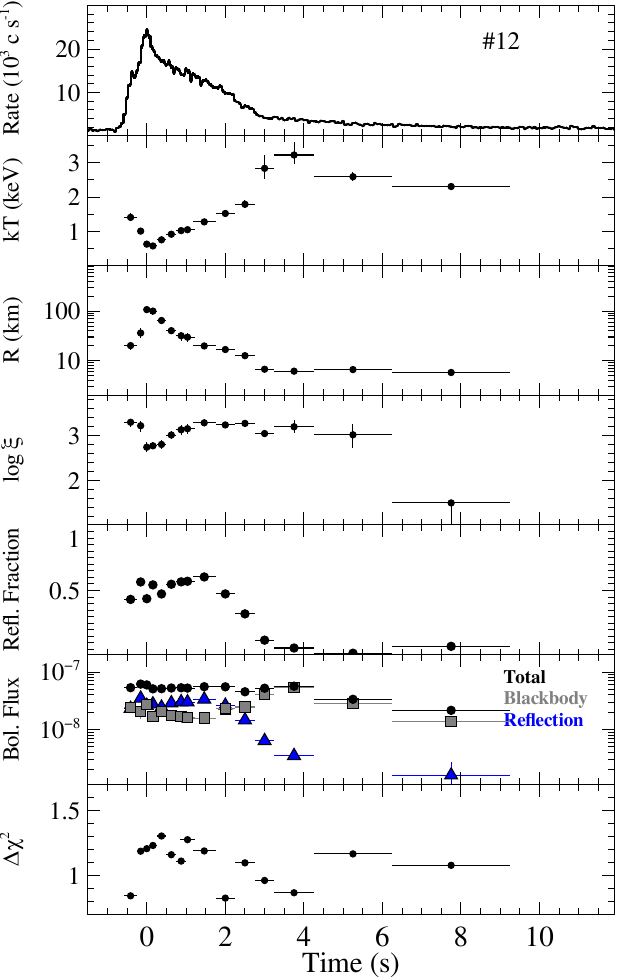}}
    \caption{Spectral parameters obtained from time-resolved spectroscopy of three representative bursts from \source using a combination of blackbody and reflection model at a fixed pre-burst emission. The black circle, grey square, and blue triangle in the sixth panel indicate the total model flux, blackbody flux, and reflection flux in the 0.1--100~keV range, respectively. }
    \label{fig:burst_spec_ref}
\end{figure*}

\subsection{Burst time-resolved spectroscopy using a reflection model} \label{sec:ref-spec}

We applied a physical reflection-based model to probe the changes in the inner accretion disk observed during strong thermonuclear bursts. In our analysis, each time-resolved spectrum contained more than 3000 counts in the 0.4--10~keV range. Instead of the variable persistent emission model, a newer version of the reflection model ({\tt bbrefl21}\footnote{\url{https://heasarc.gsfc.nasa.gov/xanadu/xspec/models/bbrefl.html}} in {\tt XSPEC}) is used to account for the soft excess emission observed from all the 15 bursts. This model is based on the works of \citet{Ballantyne2004MNRAS.351...57B} \& \citet{Ballantyne2004ApJ...602L.105B}. It is calculated considering a one-dimensional slab of uniform gas illuminated by a blackbody continuum, representing the interaction of burst photons with the inner accretion disk. The model assumes a constant disk number density of 10$^{20}$~cm$^{-3}$ with pure He abundance. The illuminating blackbody temperature and disk ionization are parameterized for a range between 0.2--4.0 keV and log~$\xi$ = 1.55 -- 4.0, respectively. Here, $\xi$ is defined as the ratio of the incident flux to helium number density in the disk and is expressed in units of erg~s$^{-1}$~cm. 

While fitting the burst spectra, we fixed the column density to the value inferred from the pre-burst emission given in Table~\ref{tab:pre-burst}. Our model also considered the contribution of pre-burst emission components that were fixed during the fitting. Moreover, we tied the temperature of the burst blackbody component with the incident blackbody temperature of the reflection model. Figure~\ref{fig:burst_spec_ref} shows burst spectral parameters evolution using the reflection model only for three bursts, such as Bursts \#3, \#7, and \#12 for representation purposes. The errors on spectral parameters are calculated for a 68\% confidence interval. This confidence range is chosen to avoid the lower or upper boundaries of reflection model parameters. In all the bursts, we observed a high value of ionization at the burst onset that decreases at the peak and recovers during the burst cooling tail. The lowest value of ionization is found after the burst (Figure~\ref{fig:burst_spec_ref}). 
The evolution of total model flux, blackbody burst flux, and flux from reflection components in the 0.1--100 keV range are also shown in the figure. The maximum value of blackbody flux is observed near the touchdown. However, the reflection appears to dominate the main burst emission between PRE peak and touchdown. This can also be seen from the variation of the reflection fraction parameter, defined as a ratio of reflection flux to total model flux (fifth panel of the figure).   

Furthermore, it is important to note that the chi-square increases around the PRE peak in both \fa and reflection models in our time-resolved spectroscopy (Figures~\ref{fig:burst_spec1} \& \ref{fig:burst_spec_ref}). The addition of a power-law with a photon index of 2--3 may well describe the spectra at this stage. The existence of power-law emission around the PRE peak signifies the cooling down of the burst where the corona becomes temporarily visible. 
We also remark that some of the residuals come from the existence of narrow spectral features in these bursts \citep{Strohmayer2019ApJ...878L..27S}. The aspects of narrow features will be presented in a separate study.


\begin{figure*}
\centering
		{\includegraphics[scale=0.51]{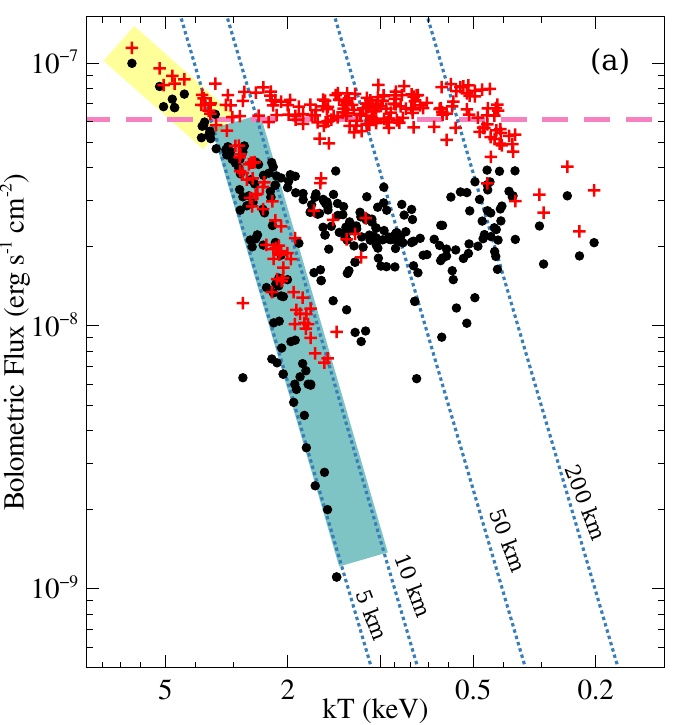}}
		{\includegraphics[scale=0.51]{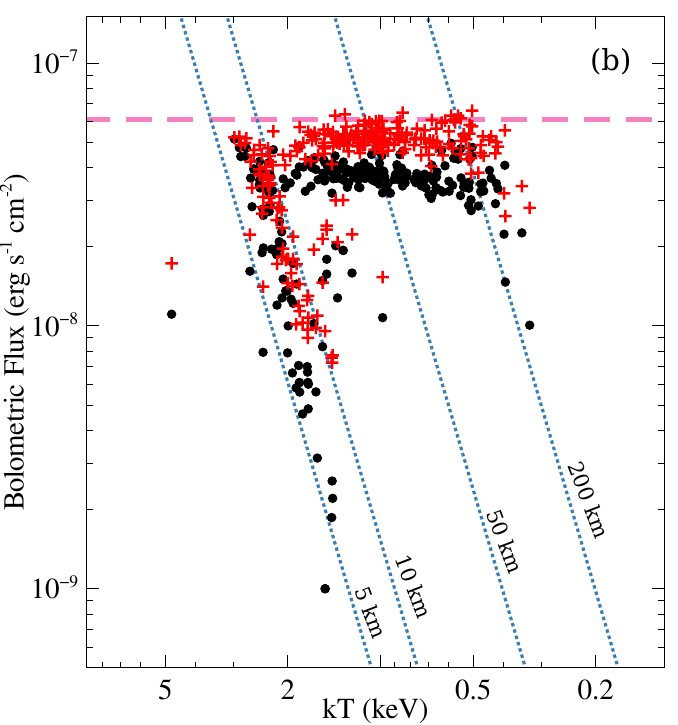}}
		{\includegraphics[scale=0.51]{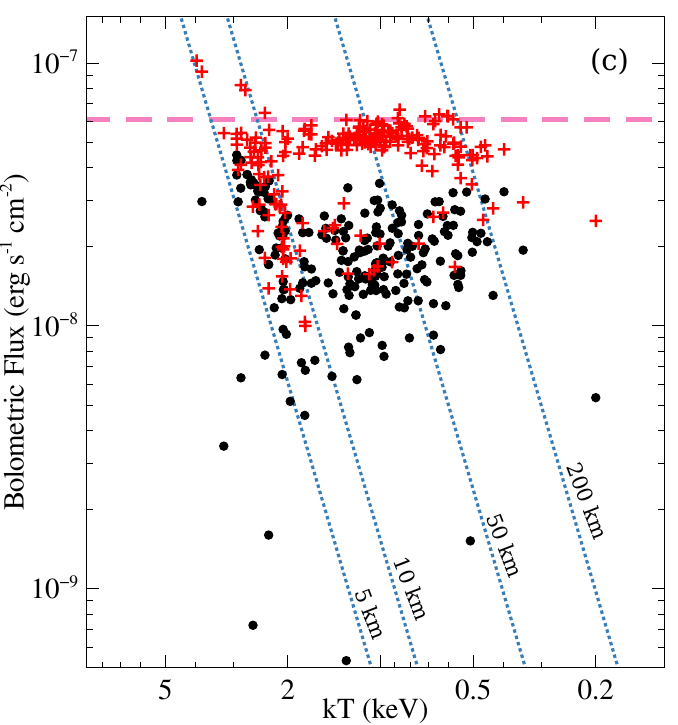}}
    \caption{Flux-temperature diagram from time-resolved spectroscopy of all 15 X-ray bursts of \source. The left (a), middle (b), and right (c) panels are from time-resolved spectroscopy of bursts using three different spectral approaches, such as the variable persistent emission model, variable pre-burst component method, and reflection model, respectively. Bolometric blackbody flux is shown with black solid dots, where the red plus symbol represents total model flux. The constant radius contours are also marked by dashed diagonal lines, assuming a distance of 8.4~kpc. The radius contours are not color-corrected (see, e.g., \citealt{2013MNRAS.429.3266G, Suleimanov2017MNRAS.472.3905S}). The horizontal dotted line corresponds to the PRE flux level.}
    \label{fig:flux-temp}
\end{figure*}

\subsection{Flux-temperature diagram} \label{sec:flux-spec}

Figure~\ref{fig:flux-temp} shows the flux-temperature diagram from the time-resolved spectroscopy including all segments of the 15 X-ray bursts of \source, and using our three different modeling approaches (panels~a, b, and c of the figure). The red and black points in each panel correspond to the total model and blackbody fluxes in the 0.1--100~keV range, respectively. The diagonal dotted lines represent constant radius contours at 5, 10, 50, and 200~km, assuming a distance of 8.4~kpc. The blackbody and total fluxes from all the bursts clearly show two distinct tracks (diagonal and horizontal) in the flux-temperature diagrams. A majority of black and red points, in the diagonal track between 5 and 10 km radius lines (see e.g., gray band in panel~a), are from the burst cooling tails. This shows that the burst cooling tails explicitly follow the standard Stefan-Boltzmann law, where the flux is proportional to the fourth power of the temperature. 
The red points in the horizontal track at radii larger than 10~km indicate the PRE nature of the bursts.  

A high blackbody temperature is usually observed at the touchdown point when the NS photosphere returns back to the surface after the PRE phase. By comparing burst parameters among the three models, we find a hotter blackbody temperature of 3 to 5~keV at the touchdown from the variable persistent emission method (panel a of Figure~\ref{fig:flux-temp}) with a minimum touchdown radius of 4--6~km from all the bursts. However, the touchdown temperature from the other two methods (middle and right panels of the figure) is below 3 keV, with a touchdown radius larger than 6~km. Interestingly, from the variable persistent emission method, the touchdown fluxes exceed the Eddington limit and follow a different slope with an index of 1.1 (yellow box in panel~a of the figure) that deviates from the standard $F \propto T^4$ relation, complicating the estimation of mass and radius of NS using the cooling-tail method (see also \citealt{2013MNRAS.429.3266G} and \citealt{Kajava2014MNRAS.445.4218K} for \rxte burst samples from \source).  This kink is not visible in the middle and right panels of Figure~\ref{fig:flux-temp}. It appears that the touchdown temperature depends on the burst modeling approach that may give rise to a kink in the burst flux-temperature diagram. 

We also found that the maximum expansion radius obtained with the  variable persistent emission method is generally 10 to 25\% larger than the values with the other two methods. The difference becomes more for (very) strong expansion bursts, possibly due to spectral degeneracy among thermal parameters in the soft X-ray band below 1~keV.  The burst fluence of each burst from the first two methods well matches within the error bars (see Table~1 for burst fluence using variable persistent emission method). The burst fluence from the reflection-based model is lower than the value from the variable persistent emission method. This is expected as the model was able to fit a limited number of spectral segments after the touchdown due to the limitation of grid parameters (section~\ref{sec:ref-spec}).

\section{Discussion} \label{sec:dis}      

We studied a sample of 15 thermonuclear X-ray bursts of \source observed in five years of \nicer data collected between 2017 and 2022. These bursts were detected during a low-hard state, corresponding to the island branch in the color-color diagram. A double-peaked profile was seen at the peak of these bursts in the 3--10~keV energy range. This behavior is mostly evident in the case of PRE bursts where the evolving photospheric temperature may lie beyond the energy bandpass of X-ray instruments. Among the 15 bursts, the 3--10~keV light curve of Burst~\#8 (see Figures~\ref{burst_lc}) resembles more like a precursor event before the burst in the absence of soft X-ray coverage. Such a profile has been seen with hard X-ray observatories like \rxte during a strong PRE burst from SAX~J1808.4--3658 \citep{2007ApJ...656..414B}.

\subsection{4U 1820--30: the fastest spinning nuclear-powered millisecond pulsar?}
Burst oscillations are fast high-frequency oscillations observed typically in a range of a few hundred Hz, with a fractional amplitude of a few percent, during X-ray bursts. These oscillations are thought to originate from bright patches of thermonuclear explosions on the NS surface. They can be detected during the burst rise, decay, or throughout the burst. The spin frequency of NSs matches closely with these oscillations \citep{Chakrabarty2003Natur.424...42C}. Mostly nuclear-powered millisecond X-ray pulsars display burst oscillations in a range of 245 to 620~Hz, with the exception of 11~Hz oscillations from IGR~J17480--2446 \citep{A.Watts2012, Bhattacharyya2022ASSL..465..125B}. The oscillation frequencies sometimes can differ by several Hz from the independent measure of NS spin frequency in millisecond pulsars.  It is still a matter of study why these oscillations do not appear in all bursts of a given source or from every NS burster in LMXBs. Currently, several theoretical models attempt to explain the observed properties of the burst oscillations. These models consider the effect of flame spreading \citep{Strohmayer1997ApJ...487L..77S}, large-scale magneto-hydrodynamical oscillations due to spreading flame \citep{Heyl2004ApJ...600..939H}, or through cooling wakes that is asymmetric emission observed during the cooling tail of the burst from the NS surface \citep{Mahmoodifar2016ApJ...818...93M}. 
Our timing analysis of burst emissions resulted in the detection of high-frequency oscillations at 716~Hz in Burst~\#11 of \source. The signal was found at a significance level of 2.9$\sigma$ using Monte Carlo simulations, with a fractional rms amplitude of 5.8$\pm$0.7\% in the 3--10~keV range. The detection of this burst oscillation at 716~Hz signifies the potential measurement of the spin frequency of the neutron star from \source for the first time.
In the case of accreting millisecond X-ray pulsars, burst oscillations are known to generally asymptotically approach the spin frequency of the pulsar as the bursts proceed \cite[see e.g.][]{A.Watts2012,Bilous2019}. The frequency evolution of the detected signal can usually be explained by an exponential function and only spans a narrow frequency range of a few Hz \cite[see e.g.][]{1999ApJ...516L..81S, 2002ApJ...580.1048M, Strohmayer2006csxs.book..113S}. We do not see any apparent frequency evolution of 716~Hz observed from Burst~\#11.

Among accreting neutron stars in LMXBs, 4U~1608--52 has shown burst oscillations at 620~Hz, which is considered to be the highest with firm detection  \citep{Hartman2003HEAD....7.1738H, Galloway2008ApJS..179..360G}. A much higher oscillation frequency was tentatively reported at 1122~Hz from XTE~J1739-285 \citep{Kaaret2007ApJ...657L..97K}. Later studies did not confirm the finding even for the same burst using independent time windows \citep{Galloway2008ApJS..179..360G}. Moreover, recent \nicer \citep{Bult2021ApJ...907...79B} and {AstroSat}  \citep{Beri2023MNRAS.521.5904B} studies of XTE~J1739-285 found prominent burst oscillation at 386.5~Hz, instead of any signal around 1122~Hz. To date, there is no known accretion or nuclear-powered neutron star that exhibits firm pulsations/oscillations above 620~Hz. If confirmed, the detection of a 716 Hz oscillation frequency from \source would make it the fastest accreting neutron star in known X-ray binary systems.

\begin{figure*}
\centering
		{\includegraphics[scale=0.55]{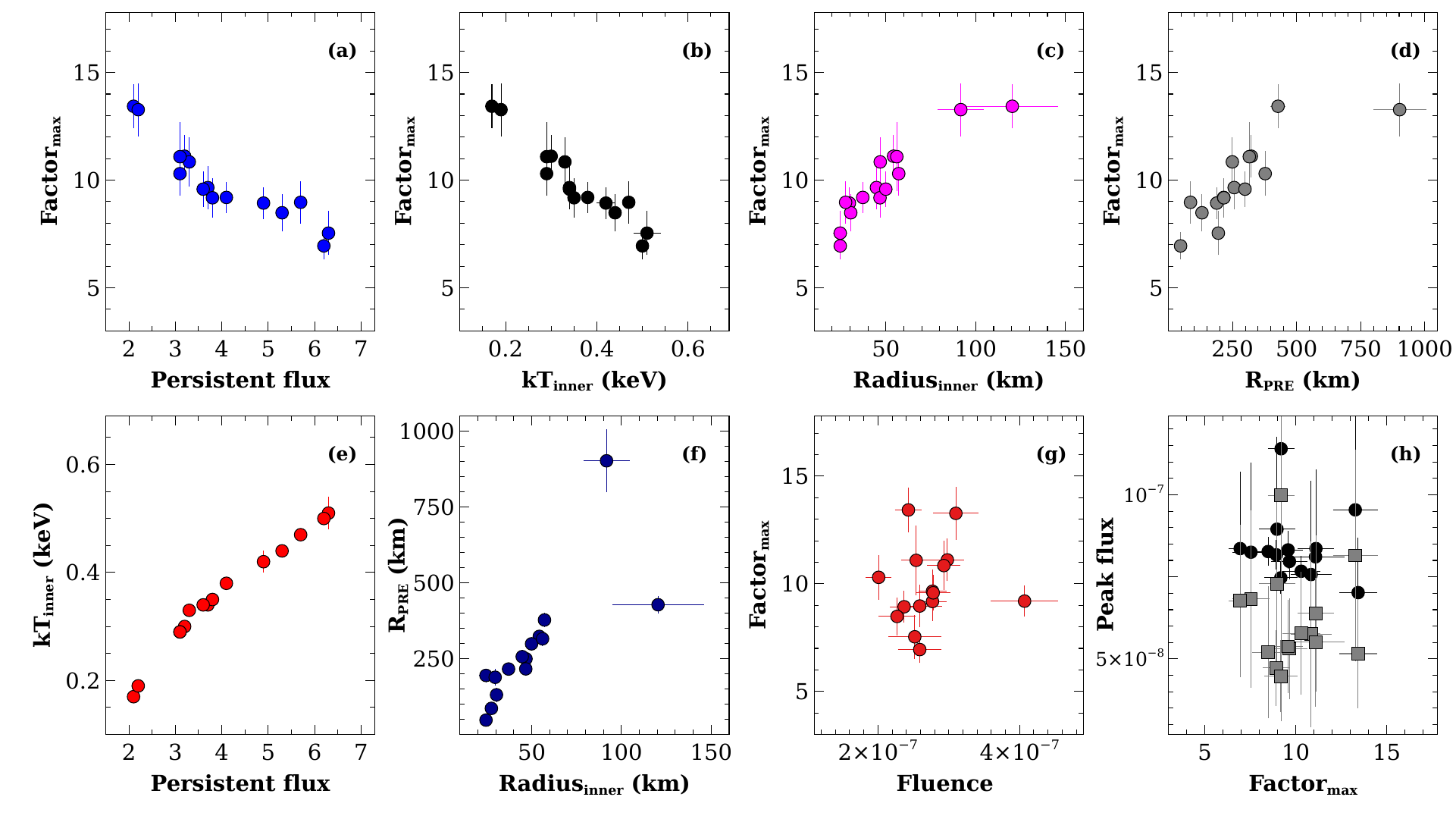}}
    \caption{ Relation between various burst and persistent spectral parameters. Unabsorbed persistent flux in 0.5-10 keV in a unit of 10$^{-9}$~erg~s~cm$^{-2}$. Factor$_{\rm max}$: maximum value of scaling factor (\fa) observed from a burst. Burst fluence: the total sum of blackbody flux integrated over each time interval from time-resolved spectroscopy in a unit of erg~cm$^{-2}$.  Peak flux: maximum value of total (black circle) and blackbody (grey square) fluxes observed from each burst in erg~s~cm$^{-2}$, respectively. The pre-burst blackbody temperature and radius (in km) are denoted with kT${\rm _{inner}}$ and Radius${\rm_{inner}}$, respectively. R${\rm_{PRE}}$ corresponds to the maximum radius attained during these PRE bursts. }
    \label{fig:burst_para_rel}
\end{figure*}

Considering non-accreting NSs, the radio pulsar PSR~J1748--2446ad in the globular cluster Terzan~5 holds the record for the fastest spinning neutron at a frequency of 716 Hz \citep{Hessels2006Sci...311.1901H}. The second fastest rotating neutron star is PSR~J0952--0607, which is a radio pulsar at a frequency of 707~Hz in the Galactic field \citep{Bassa2017ApJ...846L..20B, Nieder2019ApJ...883...42N}. 
The maximum frequency cutoff of NSs is speculated to be 730~Hz generally based on studying the spin period distribution of both nuclear- and accretion-powered millisecond X-ray pulsars   \citep{Chakrabarty2008AIPC.1068...67C, Patruno2010ApJ...722..909P}. This limit is well below the breakup spin rates (mass-shedding limit) for most equation of state models of rapidly rotating NSs \citep{Cook1994ApJ...424..823C, Chakrabarty2008AIPC.1068...67C}.

\subsection{Burst-disk interaction}

We studied the spectral properties of 15 thermonuclear X-ray bursts observed with \nicer from \source. Our time-resolved spectroscopy revealed that all these are PRE bursts with an expansion radius of more than 50~km. Moreover, a super-expansion was observed in Burst~\#8 where the NS photosphere expanded for 902$\pm$103~km with an effective temperature of 0.2 keV at the burst peak, using the variable persistent emission method. The touchdown blackbody flux from these bursts was found in a range of (4.5--10)$\times$10$^{-8}$~erg~s$^{-1}$~cm$^{-2}$ that is well in agreement with the Eddington limited flux as per MINBAR studies \citep{Galloway2020ApJS..249...32G}. 
Although our spectroscopy results are broadly consistent with the study performed using the variable persistent emission method by \citet{Yu2023arXiv231216420Y}, some differences in blackbody temperature and radius parameters are found. This is likely due to the choice of different time durations and the number of segments considered for the time-resolved spectroscopy of X-ray bursts in our study. Moreover, \citet{Yu2023arXiv231216420Y} considered a 64~s exposure of pre-burst emission for all the bursts, unlike a relatively longer exposure used in the present study (section~\ref{sec:per-spec}). This also affects pre-burst parameters, leading to a difference in \fa and other parameters.   We discuss some of the additional findings from our studies in this section.   

Based on the variable persistent emission model, the scaling factor \fa varied during these bursts by more than an order of magnitude at the burst peak (Figure~\ref{fig:burst_spec1}). This suggests that the (pre-burst) accretion emission evolves effectively during the bursts, altering the persistent accretion geometry. Such behavior can be understood in terms of changes in accretion mass rate due to intense radiation drag (Poynting-Robertson effect) that could drain out the inner accretion disk onto the NS \citep{Walker1992, Degenaar2018}. A reprocessing or reflection of burst photons with the accretion disk can also be speculated at this stage. 
From Figure~\ref{fig:burst_spec1}, we also note that the observed maximum \fa peaks a couple of seconds earlier than the maximum burst flux at the touchdown. This time delay may signify that the pre-burst accretion gets altered even before the maximum burst emission as suggested by \citet{2018ApJ...867L..28F, Fragile2020}.

To examine the connection between X-ray bursts and persistent emission, we studied the evolution of various bursts and accretion parameters in Figure~\ref{fig:burst_para_rel}. Our findings are listed below:  

\begin{enumerate}
\item Panel~(a) of the figure shows the evolution of the maximum value of scaling factor (Factor${\rm _{max}}$) observed from each burst with its corresponding (pre-burst) persistent flux. These parameters exhibit a negative correlation.  This behavior is consistent with the relationship observed between Factor${\rm _{max}}$ and the pre-burst accretion rate of all PRE bursts from different sources with \rxte in Figure~10 of  \citet{Worpel2013}. 

\item From Panel~(b) of the figure, Factor${\rm _{max}}$ follows a negative correlation with the inner disk temperature. This is obvious as persistent flux is positively dependent on the inner disk temperature, as shown in Panel~(e).   

\item From Panel~(c), Factor${\rm _{max}}$ is positively correlated with the inner disk radius. 

\item The Factor${\rm _{max}}$ seems to be correlated with the maximum photospheric radius reached during a burst (Panel~d).  
Considering the fact that the $fa$ parameter represents the accumulative changes in pre-burst emission during the burst and can not be easily disentangled between separate components of the persistent emission, we suggest caution in a direct physical interpretation of relations as found in panels~(a)-(d) of Figure~\ref{fig:burst_para_rel}.

\item Since all these are PRE bursts, we do not observe any correlation between Factor${\rm _{max}}$ and peak burst flux (grey) in Panel~(h). This is an interesting characteristic of PRE bursts. For example, in the case of non-PRE bursts from 4U 1636–536 \citep{Guver2022ApJ...935..154G}, a positive correlation between these parameters was observed. 

\item Additionally, we found that burst peak flux, total bolometric flux, as well as burst fluence, are almost constant with the persistent flux for all these Eddington limited bursts. These results are not shown in Figure~\ref{fig:burst_para_rel}.

\item A positive relation is seen between the inner disk radius and the maximum photospheric radius reached during a burst (panel~f of Figure~\ref{fig:burst_para_rel}). 
This may hint that when the inner accretion disk is far away from the neutron star, the photosphere expands more easily during the PRE bursts. 
Such expansion allows the burst photosphere to interact closely with the persistent environment, including the inner and outer parts of the accretion disk. A similar behavior was anticipated for super-expansion bursts with \rxte where the expansions were found at the lowest mass accretion rate \citep{Zand2012A&A...547A..47I}. 

\end{enumerate}

One of the interesting aspects we observed in Figure~\ref{fig:burst_para_rel} is the correlation between the inner disk and maximum photospheric expansion radii. It appears the accretion geometry holds control over the free expansion of PRE bursts up to a certain degree. The burst photosphere evolves more freely to larger distances when the accretion disk is farther away from the NS surface. Due to the presence of an (optically thick) accretion disk and (optically thin) coronae close to the surface, the photosphere could not expand easily to larger distances.  However, such an interpretation needs to be considered with caution due to limited (theoretical) simulation studies of the interaction between accretion disk and strong PRE bursts.

Recently, \citet{Yu2023arXiv231216420Y} studied these \nicer bursts using the {\tt relxill} reflection model, in addition to the variable persistent emission method. Our results with the $fa$-method are mostly consistent with their study. The authors have also reported the presence of a super-expansion of 1318~km in the case of Burst~\#8. The difference in values of maximum expansion comes from the choice of spectral time segments in our study.  
Following the {\tt relxill} reflection, \citet{Yu2023arXiv231216420Y} found a large variation in the disk density between 10$^{15}$ and 10$^{19}$~cm$^{-3}$ reported during Burst~\#7 and \#8. Such a strong change in disk density seems unfeasible at a shorter time scale of the burst but could signify the changes in the accretion disk.  In our study, we assumed a constant disk density of 10$^{20}$~cm$^{-3}$ in {\tt bbrefl21}, with a helium abundance that is suitable to the accreted matter composition in the system (see section~\ref{sec:ref-spec}). The approach allowed us to probe the change in the inner disk ionization parameter across the burst.  

Based on our reflection model, we examined the changes in the inner disk ionization which dips at the burst peak before reaching its maxima around touchdown (Figure~\ref{fig:burst_spec_ref}).  The observed behavior can be understood in the following manner. The NS photosphere evolves rapidly within a few seconds time scale for these PRE bursts. Initially, the burst photons affect the accretion flow in the inner accretion disk, resulting in a high ionization at the onset. As the photosphere evolves to larger radii, the inner part of the disk perhaps loses its capability for ionization because of disturbed flow, resulting in a slight drop in the ionization parameter near the burst peak. The drop in ionization may also come from the outer part of the disk, which is weakly ionized at the PRE phase.  When the photosphere retreats towards the surface, the inner accretion disk again contributes to ionization. After the burst touchdown,  a low value of ionization close to the model's lower boundary is evident.
The burst spectroscopy using the reflection model also suggests that the inner accretion disk is getting altered due to the NS photosphere evolution.  At large photosphere expansion, the inner disk does not contribute much due to its disturbed accretion flow.

\section{Conclusion} \label{sec:conclusion}
We have studied thermonuclear bursts observed from \source in its low-hard state using \nicer. Based on time-resolved spectroscopy, we confirm that these bursts are of PRE nature, where the photosphere expanded for more than 50 km. In the case of Burst~\#8, the super-expansion was measured up to 902~km. We observed a negative correlation between persistent emission and the maximum \fa value from these bursts.  Following the reflection model, we also probed the changes in the inner-accretion disk through the evolution of the ionization parameter. The immense radiations from thermonuclear bursts affect the persistent geometry, even altering the inner disk flow. 

We also compared the time-resolved burst spectroscopy parameters from all three models using the flux-temperature diagram. The touchdown temperature obtained from the variable persistent emission model is usually much hotter than that from the other two models. This leads to a kink with a slope of 1.1 in the diagram around the touchdown, deviating from the standard Stefan-Boltzmann law where the flux is proportional to the fourth power of temperature. This deviation is not observed in the other two models. The presence of two distinct slopes, one at the touchdown and another during the cooling tails of bursts, suggests that the variable persistent emission model may not adequately describe the parameters around the touchdown of these PRE bursts from \source. This emphasizes the need for additional modeling approaches, such as the reflection model or alternative method presented in the paper.

Our timing analysis led to the detection of candidate burst oscillations at a frequency of 716~Hz throughout Burst~\#11 in the 3--10 keV range. The oscillation signal was detected at 2.9$\sigma$ significance level based on Monte Carlo simulations.  Assuming the observed frequency is the NS spin rotation, \source can become the fastest-rotating NS discovered among X-ray binaries. The source is also on par with the fastest rotating NS, i.e., radio pulsar PSR~J1748--2446 with a spin frequency of 716~Hz. This finding is interesting as the frequency cutoff limit of NS is expected to be 730~Hz as per Bayesian analysis of observable accretion- and nuclear-powered pulsars. Our finding is also in line with the lack of the sub-millisecond pulsar above 730~Hz.

\section*{Acknowledgements}
We thank the reviewer for constructive suggestions that improved the manuscript. This work was supported by NASA through the \nicer mission and the Astrophysics Explorers Program and made use of data and software provided by the High Energy Astrophysics Science Archive Research Center (HEASARC). 
ZFB, TB, and TG have been supported by the Scientific Research Projects Coordination Unit of Istanbul University (ADEP Project No: FBA-2023-39409), in part by the Turkish Republic, Presidency of Strategy and Budget project, 2016K121370, and in part by the Scientific and Technological Research Council (T\"UB\.ITAK) 119F082.

\section*{ Data Availability}
All the data used in this paper are publicly available
through NASA/HEASARC archives.  The data products and other files will be made available on request.

\bibliography{references}{}
\bibliographystyle{aasjournal}

\end{document}